\documentclass[pre,aps,floatfix,preprint]{revtex4-1}

\usepackage{amsmath,amssymb}
\usepackage{graphicx}
\usepackage{psfrag}
\usepackage{graphicx}
\usepackage{epsf}

\def\beq{\begin{equation}}
\def\eeq{\end{equation}}
\def\bea{\begin{eqnarray}}
\def\eea{\end{eqnarray}}

\begin{document}

\title{Fluctuations and symmetries in two-dimensional active gels}
\author{Niladri Sarkar and Abhik Basu} \affiliation{{Theoretical Condensed Matter
Physics Division, Saha Institute of Nuclear Physics, 1/AF,
Bidhannagar, Kolkata (Calcutta) 700 064, India} }

\date{\today}

\begin{abstract}
Motivated by the unique physical properties of {\em biological active matter}, e.g., cytoskeletal dynamics in eukaryotic cells, we set up  {\em effective} two-dimensional ($2d$) coarse-grained hydrodynamic equations for the dynamics of thin {\em active gels} with polar or nematic symmetries. We use the well-known three-dimensional ($3d$) descriptions  [K. Kruse {\em et al}, {\em Eur. Phys. J E}, {\bf 16}, 5 (2005); A. Basu {\em et al}, {\em Eur. Phys. J E}, {\bf 27}, 149 (2008)] for thin active gel samples confined between parallel plates with appropriate boundary conditions to derive the effective $2d$ constitutive relations between appropriate thermodynamic fluxes and generalised forces for small deviations from equilibrium. We consider three distinct cases, characterised by spatial symmetries and boundary conditions, and show how such considerations dictate the structure of the constitutive relations. We use these to study the linear instabilities, calculate the correlation functions and the diffusion constant of a small tagged particle, and elucidate their dependences on the {\em activity} or nonequilibrium drive.
\end{abstract}

\maketitle

\section{Introduction}

The dynamics of equilibrium systems describe time evolutions of fluctuations around the minimum free energy or maximum entropy states. Systems in equilibrium do not consume energy continuously. Dynamics in such systems are characterised by the Fluctuation-Dissipation-Theorem (FDT), relating a susceptibility with an appropriate correlation function \cite{fdt}. In contrast, {\em active} systems are driven out of equilibrium by a continuous consumption (supply) of energy. Well-known examples are gels driven by chemical reactions \cite{exm1}, vibrating granular materials \cite{exm2}, large scale coordinated motion of self-propelled particles like bacteria colonies and bird flocks \cite {exm3} and cell cytoskeleton \cite{alberts}. The physically interesting aspect of biologically relevant active matter is its ability to convert free energy (available in the form of chemical energy, e.g., hydrolysing Adenosine-Triphosphate (ATP)) into mechanical work and systematic movement. Many of these systems, despite having very different characteristic length and time scales, and microscopic details, share common general features in the long wavelength, large time limit where conservation laws and symmetries (and not microscopic details) govern the general behaviour. Such generalities and hence coarse-grained approaches based on them are therefore generic. Coarse-grained descriptions in statistical physics have a long  and successful history, beginning with equilibrium critical phenomena \cite{crit} and equilibrium critical dynamics \cite{crit-dyn}, and more recently, to a variety of systems out of equilibrium \cite{simha,hatwalne}. Of late, such coarse-grained approach has been applied extensively to understand the physics of biologically motivated systems, see, e.g., instabilities of cortical actin layer \cite{cortical}. Nonequilibrium nature of the fluctuations in cell, and hence violation of the FDT,
has been tested experimentally by Mizuno {\em et al} \cite{mizuno}.
Hydrodynamic theories have been developed which successfully describes some dynamic cellular processes \cite{hydro}. Further, formation of specific patterns in cytoskeletal structures is common, e.g., bundles, asters, vortices etc. \cite{pattern}. Ref.~\cite{alex} provides a coarse-grained theory at the mesoscopic scale for cortical patterns in plant cells. For further details, we refer the reader Refs.~\cite{menon,sriram-condmat,seminar} for recent reviews on these subjects.

In this paper we focus on the dynamical behaviour of a cortical layer of filaments under various circumstances (e.g., various spatial symmetries, boundary conditions) at length scales much larger than the filament lengths and layer thickness with polar as well as nematic macroscopic ordering, for which a generic coarse-grained continuum $2d$ description would be appropriate. Our main achievement is a set of coarse-grained hydrodynamic equations of motion for the orientational order parameter, local density and the velocity fields in the frictional limit, which we use to obtain results on the macroscopic properties of the underlying systems subject to various conditions. We consider the three following distinct cases for a thin active gel:
(i) In-plane nematic or polar order without any external force (hereafter {\bf System I}, (ii) No in-plane order (ordering is normal to the plane) (hereafter {\bf System II} and (iii) In-plane polar order with external (surface) forces (hereafter {\bf System III}).
In each of the cases, we set up the {\em effective} $2d$ equations by using the $3d$ framework of Refs.~\cite{jfjoanny,abmpi} for $3d$ active gels and averaging them over the thin direction. We show that such effective $2d$ equations are consistent with the macroscopic symmetries of the systems. We show that our equations for finite {\em activity} (characertised by a constant parameter $\Delta\mu$ in our notation) break time-reversal invariance explicitly and hence display nonequilibrium behaviour. In each of the cases, as the magnitude of $\Delta\mu<0$ exceeds a critical value $\Delta\mu_c$, the initial chosen states exhibit finite wavevector instability, leading to anisotropic patterned states. Below the threshold of instability, i.e., $|\Delta\mu| <|\Delta\mu_c|$, equal-time density correlations function display giant fluctuations in some of the cases considered by us, as also reported previously in Refs.~\cite{simha,toner} for active nematics and polar flocks. In addition, we find generic underdamped propagating modes for systems with in-plane polar order both below and above the threshold of linear instability.
Our fluctuating hydrodynamic theory of $2d$ active matters in contact with a substrate should be well-suited to describe a thin actin cytoskeleton in contact with the bulk of the cell, which serves as a substrate. Refs.~\cite{jfjoanny,joanny2} briefly discussed dynamical properties of a quasi-$2d$ active gel layer in contact with a substrate.  Here, we extend their work and provide systematic studies of quasi-$2d$ thin active gel layers in contact with substrates together with various boundary conditions and initial reference states. We show how effective friction, which dominates the velocity field dynamics, emerges in each of the cases out of the lubrication approximation \cite{lub} we make here. In addition, the two-dimensional description of actin dynamics in a motility assay is related to one of the cases considered by us here.
The problems which we address here broadly refer to the phenomenon of {\em flocking}: the collective motion of self-propelled active particles; see Ref.~\cite{sriram-toner} for a review of recent results. We use the framework developed in Ref.~\cite{jfjoanny} (equivalently, Ref.~\cite{simha}). The remaining part of this paper is structured as follows: In Sec.~\ref{symm} we discuss the appropriate $2d$ thermodynamic fluxes and conjugate generalised forces for this problem. In Sec.~\ref{3dto2d}, we derived the constitutive relations between them. Then in Sec.~\ref{timerev} we add FDT-obeying thermal noises in the $2d$ equations when $\Delta\mu=0$. In Sec.~\ref{eom}, we use our constitutive relations to illustrate linear instabilities at finite wavevector and calculate correlation functions of the local polarisation and density variables. We further calculate the diffusion coefficient of a tagged particle and highlight its non-trivial dependence on the activity. In Sec.~\ref{conclu} we summarise and conclude. Finally, a short comment on the notations used in the article is in order: We use a Roman subscript (e.g., $i$) to denote components of a $2d$ vector, i.e., $i=x,y$ and a Greek subscript (e.g., $\alpha$) to denote the components of a $3d$ vector, i.e., $\alpha = x,y,z$.

\section{Genaralised fluxes and forces}
\label{symm}

We develop a dynamical theory of active gels in the coarse-grained (continuum) hydrodynamic approach. We consider here a single fluid model where the embedding solvent is at rest \cite{abmpi}. The system is assumed to be only slightly away from equilibrium. This available free energy due to the active, nonequilibrium processes is equivalent to a nonequilibrium generalised force, which breaks the time reversal symmetry,  represented by $\Delta\mu$ in our description. 
The linear constitutive relations between the appropriate fluxes and generalised forces, which we obtain as
a general expansion of the fluxes in terms of the forces in the spirit of Onsager reciprocity relations for fluctuations in equilibrium, holds for active gels close to thermodynamic equilibrium. The detail form of the linear expansion of the fluxes in terms of the forces depend upon the symmetries of the system under consideration (see below).
We consider the dynamics in the high friction limit. Hence, the constitutive equations we set up are valid {\em only in a preferred frame of reference} and as a result there is no Galilean invariance of the system - they will not be invariant under the transformation ${\bf v}\rightarrow {\bf v} + {\bf v}_0$. Here ${\bf v}$ is the local velocity of actin gel filaments. Further since we are modeling a gel with local nematic or polar order, we introduce a polarisation field $\bf p$, a vector, to describe the local orientation of the filaments making up the gel. With every filament one can associate a unit vector pointing to one end. The vector $\bf p$ is given by the local average of a large number of these unit vectors. Our choice of flux is $(v_i, \dot{p_i}),\;i=x,y$.  The field $\dot{{\bf p}}$ is the time derivative of the filament orientational field ${\bf p}$. Our choice for fluxes reflect the lack of Galilean invariance in the system. The corresponding generalised forces are $(F_i, h_i)$ such that the rate of change in the Free energy $\mathcal F$ per unit surface may be written as
\begin{equation}
\dot {\mathcal F}=-\int d^2r [-{v_i \partial_j\sigma^d_{ij}} + h_{\perp i}\dot p_i]=-\int d^2r [-v_i\nabla_i \Pi +h_i \dot p_i]=-\int d^2r [v_i F_i + h_i \dot p_i],
\label{freenderi}
\end{equation}
where $\sigma^d_{ij}$ is the total dissipative stress and we have used the force balance condition $\partial_j \sigma^d_{ij} =\partial_i \Pi$ when there are no external forces in the system and $F_i\equiv -\nabla_i\Pi$. Note, for a pure fluid confined between two plates, the velocity $v_i$ in the high friction limit is proportional to the spatial derivative of pressure $\Pi: v_i\propto -\nabla_i\Pi$ ({\em Darcy's law}). In the present problem the Darcy's law will be generalised to include forces due to the orientation field. Further, $h_{\perp i}$ is the field thermodynamically conjugate to $\dot{p_i}$. Here, $"\perp"$ refers to the $XY$ plane. Under $t\rightarrow -t$, we have $F_i\rightarrow F_i$ and $h_i\rightarrow h_i$. The rate of change of free energy has, in general two parts - reversible $\dot{\mathcal F}_{rev}$ and irreversible $\dot{\mathcal F}_{irr}$ parts. The former is the rate of change in the internal energy and the latter is linked to the entropy production. The fluxes also, in general, can be decomposed into reactive and dissipative parts. Note that the generalised forces have well-defined signatures with respect to time reversal. The dissipative fluxes have the same signature under time reversal as their conjugate forces, while reactive fluxes have opposite signatures under time reversal. In the present problem,  both the generalised forces are even under time-reversal, and, therefore, the reactive fluxes vanish. Thus, according to the Onsager reciprocity theorem \cite{onsager}, the Onsager matrix will be fully symmetric in the viscous limit. Spatial symmetries of the system will dictate the detail form of the Onsager matrix. We will see below in Sec.~\ref{3dto2d} that our derivation of the $2d$ constitutive relations from the $3d$ constitutive relations \cite{joanny2,abmpi} are fully consistent with the discussions in the Section.

\section{Effective $2d$ descriptions of thin confined $3d$ samples}
\label{3dto2d}

In this Sec. we show how two-dimensional descriptions emerge naturally for thin confined three-dimensional ($3d$) samples with appropriate boundary conditions under averaging over the thickness. This averages out the variation along the thin direction and leaves only the in-plane variations, leading to an {\em effective} $2d$ description, which may also be derived {\em apriori} by imposing symmetry conditions between $2d$ fluxes and forces (see Secs.~\ref{appen1} and \ref{appen2}).

Consider a thin sample of thickness $W$, confined between two parallel plates, parallel to the $XY$ plane. It is assumed to be of infinite extent along the $x$ and $y$-directions and the $z$-direction becomes confined; $W$ is considered to be much smaller than the lateral in-plane dimensions. In what follows below, we set $v_z=0$ in thin film approximation, which may be justified as follows: The incompressibility condition on the $3d$ velocity field yields
\beq
\partial_z v_z + \nabla_\perp \cdot {\bf v}_{\perp}=0,
\eeq
[${\bf v}_\perp = (v_x,v_y)$] yielding (in order of magnitude)
$v_z\sim \frac{W}{L}v_\perp$,
where $L\rightarrow \infty$ is the lateral linear size of the system and derivatives has been replaced by the corresponding inverse length scales. Thus in the limit of small $W/L$, $v_z$ is negligibly small and is set to zero. Even though the above approximation rests on the incompressibility of the fluid, we use this approximation for the compressible case as well.

We begin from the $3d$ constitutive relations and the force balance equations, as reported in Ref.~\cite{jfjoanny}. The appropriate thermodynamic fluxes are the symmetric part of the stress tensor $\sigma_{\alpha\beta}$ and local polarisation field $p_\alpha$. The conjugate generalised forces are the strain-rate tensor $u_{\alpha\beta}\equiv (\partial_\alpha v_\beta + \partial_\beta v_\alpha)/2$ and the local orienting field $h_\alpha$. Here, $\alpha,\,\beta$ are $x,y$ or $z$. The rate of change of the free energy in $3d$ $\dot{\mathcal F}_{3d} = -\int d^3 r [u_{\alpha\beta}\sigma_{\alpha\beta} + \dot p_\alpha h_\alpha] =  - \int dxdy \int_{-W/2}^{W/2} dz [u_{\alpha\beta}\sigma_{\alpha\beta} + \dot p_\alpha h_\alpha]=-\int dxdy \int_{-W/2}^{W/2} dz [u_{\alpha\beta} \sigma_{\alpha\beta}^{tot} + \dot p_\alpha h_\alpha]$, where $\sigma_{\alpha\beta}^{tot} = \sigma_{\alpha\beta} + \sigma_{\alpha\beta}^a,\,\sigma_{\alpha\beta}^a$ is the antisymmetric part of the total stress tensor $\sigma_{\alpha\beta}^{tot}$. We impose specific boundary conditions (e.g., no-slip) on the velocity fields at the confining walls: $v_\alpha(x,y,z=\pm W/2)=0$. This breaks the Galilean invariance, as the rest frame of the confining walls becomes the preferred frame of reference, in which our equations of motion will be valid. Under partial integrations, the rate of change of free energy becomes
\begin{equation}
\dot{\mathcal F}_{3d}=-\int dxdy \int_{-W/2}^{W/2} dz [-\partial_\beta\sigma_{\alpha\beta}v_\alpha + \dot p_\alpha h_\alpha].
\label{freedecay}
\end{equation}
In systems with small sizes (biological gels are of the order of micrometers), such as ours, momentum conservation is replaced by the force balance condition \cite{jfjoanny}
\beq
\partial_\beta \sigma^{\rm tot}_{\alpha\beta} -\partial_\alpha \Pi = 0,
\label{forcebal}
\eeq
in the absence of any external forces. Here, $\sigma^{\rm tot}_{\alpha\beta}$ is the total stress tensor. With the force balance condition, Eq.~(\ref{freedecay}) reduces to
\begin{equation}
\dot{\mathcal F}_{3d}=-\int dxdy  \int_{-W/2}^{W/2} dz[-\partial_\alpha\Pi v_\alpha +\dot p_\alpha h_\alpha],
\end{equation}
suggesting that the pair $(v_\alpha,\partial_\alpha \Pi)$ may be treated as a thermodynamic flux-force in a linear response theoretic description. Moreover, treating $v_\alpha$, as opposed to $u_{\alpha\beta}$ as a flux automatically breaks the invariance under a Galilean boost: ${\bf v} \rightarrow {\bf v} + {\bf v}_0$, which is expected in the presence of a wall. This is in agreement with our arguments given in Sec.\ref{symm}. Finally, averaging over the $z$-direction and expressing in terms of the mid-plane values of the fields, we obtain
\begin{equation}
\dot{\mathcal F}=-\int dxdy [-\partial_i\Pi v_i +\dot p_i h_{\perp i}],
\end{equation}
which matches with Eq.~(\ref{freenderi}). Here, $i=x,y$.

The $3d$ constitutive relations are \cite{jfjoanny}
\begin{eqnarray}
\sigma&=&2\eta_1 u - \Delta \mu (\xi p^2+\bar{\xi}) + \bar{\nu}_1p_{\alpha}h_{\alpha} \\
\tilde{\sigma}_{\alpha\beta}&=&2\eta_2 \tilde{u}_{\alpha\beta}-\xi'\Delta\mu(p_{\alpha}p_{\beta}-{p^2 \over d}\delta_{\alpha\beta})  + {\nu_1 \over 2}\{p_{\alpha}h_{\beta}+p_{\beta}h_{\alpha} -{2 \over d}\delta_{\alpha\beta}p_{\gamma}h_{\gamma}\}, \\
{Dp_{\alpha} \over Dt}&=& {h_{\alpha} \over \gamma_1}+\lambda_1 p_{\alpha} \Delta\mu - \nu_1\tilde{u}_{\alpha\beta}p_{\beta}-\bar{\nu}_1 u p_{\alpha} +\xi_A {\bf p\cdot\nabla} p_\alpha +\xi_B {\bf p_\alpha} {\nabla\cdot \bf p},
\label{ons3dp}
\end{eqnarray}
where $\sigma=\sigma_{\alpha\alpha}$ is the trace of $\sigma_{\alpha\beta}$, $\tilde\sigma_{\alpha\beta}=\sigma_{\alpha\beta} - \frac{1}{3}\sigma \delta_{\alpha\beta}$ is the traceless symmetric part of $\sigma_{\alpha\beta}$. Similarly $u$ and $\tilde u_{\alpha\beta}$ are trace and traceless symmetric parts of $u_{\alpha\beta}$, $\eta_1$ and $\eta_2$ are bulk and shear viscosities respectively, coefficients $\nu_1,\,\overline\nu_1$ couple with the orientational degrees of freedom, $\xi,\,\xi'$ and $\overline\xi$ are coupling constants which parametrise the active stress. The derivative $\frac{D}{Dt}$ is the covariant derivative and is defined by $\frac{DA_\alpha}{Dt}\equiv \frac{\partial A_\alpha}{\partial t} +{\bf v}\cdot\nabla A_\alpha + \omega_{\alpha\beta}A_\beta$ for any vector field $A_\alpha$ with $\omega_{\alpha\beta}=(\partial_\alpha v_\beta -\partial_\beta v_\alpha)/2$ as the vorticity tensor. Terms with coefficients $\xi_A$ and $\xi_B$ are polar terms which break the nematic symmetry. For a nematic sample $\xi_A=0=\xi_B$.
The antisymmetric part of the stress tensor is given by $\sigma_{\alpha\beta}^a={1 \over 2}(p_{\alpha}h_{\beta}-p_{\beta}h_{\alpha})$. Therefore the total stress becomes
\begin{eqnarray}
\sigma_{\alpha\beta}&=&\sigma\delta_{\alpha\beta}+\tilde{\sigma}_{\alpha\beta}+\sigma_{\alpha\beta}^a \nonumber \\
&=&2u\delta_{\alpha\beta}(\eta_1-{\eta_2 \over d})+\eta_2(\partial_{\beta}v_{\alpha}+\partial_{\alpha}v_{\beta}) + (\bar{\nu}_1-{\nu_1 \over d})p_{\gamma}h_{\gamma}\delta_{\alpha\beta} + {\nu_1 \over 2}(p_{\alpha}h_{\beta}+p_{\beta}h_{\alpha})\nonumber \\
&& - \Delta\mu(\xi+\bar{\xi}-{\xi' \over d})\delta_{\alpha\beta}p^2-\xi'\Delta\mu p_{\alpha}p_{\beta} +{1 \over 2}(p_{\alpha}h_{\beta}-p_{\beta}h_{\alpha}).
\end{eqnarray}

The force balance condition (\ref{forcebal}) then  yields the generalised Stokes equation
\begin{eqnarray}
\partial_{\beta}\sigma_{\alpha\beta}&=&\partial_{\alpha}\Pi  = 2(\eta_1-{\eta_2 \over d})\partial_{\alpha}u + (\bar{\nu}_1-{\nu_1 \over d})\partial_{\alpha}(p_{\gamma}h_{\gamma}) + \eta_2\nabla^2 v_{\alpha} \nonumber \\ &+& \eta_2\partial_{\alpha}\nabla\cdot{\bf v} - \xi'\partial_{\beta}(p_{\alpha}p_{\beta})\Delta\mu + {\nu_1 \over 2}\partial_{\beta}(p_{\alpha}h_{\beta}+p_{\beta}h_{\alpha}) + {1 \over 2}\partial_{\beta}(p_{\alpha}h_{\beta}-p_{\beta}h_{\alpha}). \label{genstokes}
\end{eqnarray}
We assume there are no mean flows in the system, i.e., in the steady (without fluctuations) state ${\bf v}=(0,0,0)$. In addition, we consider three different unperturbed (without fluctuations) reference states for $\bf p$ with appropriate boundary conditions. In each of the cases below, we linearise about the chosen reference states of $\bf v$ and $\bf p$. We use the Franck free energy for the nematic liquid crystals, which, in the equal Franck's constant limit, is given by
\begin{equation}
{\mathcal F}_p = \frac{1}{2}\int d^3 r \kappa (\partial_\alpha p_\beta)^2=\frac{1}{2}\int dx dy \int_{-W/2}^{W/2} dz\,\kappa (\partial_\alpha p_\beta)^2.
\label{Frank}
\end{equation}
We discard the longitudinal field (equivalently the {\em Lagrange multiplier}) as its only role is to renormalise the bare coefficients which do not change the results at the scaling level. Finally, in some of the cases discussed below we do not consider any anchoring conditions on the actin filaments. Apart from its theoretical interests, such no anchoring conditions may arise for small-size objects like actin filaments where active anchoring conditions may compete against traditional liquid crystal anchroing conditions arising due to energy or entropic reasons at the confining surfaces.

\subsection{System I: $2d$ Planer polar/nematic order without external forces}
\label{config1}

We begin with the reference unperturbed state ${\bf p_0}=(1,0,0),\;\;(v_x,v_y,v_z)=(0,0,0)$. We show below that the {\em effective} $2d$ descriptions of small fluctuations around this state is a $2d$ state with {\em planer polar/nematic order without external forces.}
Since the $x$-axis is the ordering direction, the system should be invariant under $y\rightarrow -y$ as nothing distinguishes between $+y$ and $ -y$ directions. However, there is no $x\rightarrow -x$ symmetry. Further, since $p_x=1$, the fluctuations in $p_x$ are higher order in smallness. Moreover, if the sample is nematic, it should also be invariant under ${\bf p}\rightarrow -{\bf p}$. This symmetry is however absent for a polar sample. We consider both nematic and polar samples here.
The effective $2d$ description in this case should be invariant under these symmetries, a fact which we confirm below.
\begin{figure}[h]
\includegraphics[width=9cm]{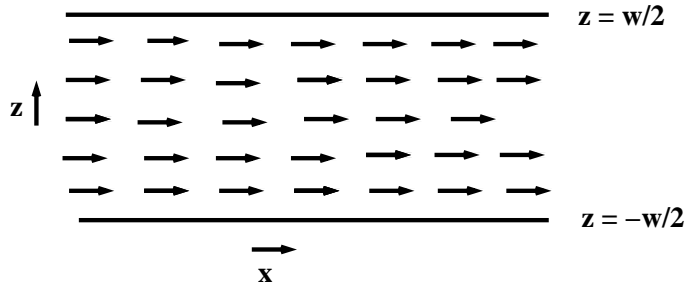}
\caption{A schematic diagram of the chosen initial reference state for System 1. Arrows (parallel to the $x$-axis) indicate the direction of
$\bf p$ in the reference state.}
\end{figure}

We impose strong anchoring boundary conditions: $p_{z}=0$ at $z=\pm W/2$ and we impose no restriction on $p_{y}$ at $z=\pm W/2$. We first consider the apolar case. Since the fluctuations in $p_{x}$ are second order in smallness we can safely ignore the dynamics of $p_{x}$. Thus the fluctuations ${\bf \delta p}= (0,p_y,p_z)$. Therefore, we consider the dynamics of $p_{y}$ and $p_{z}$ only.
Further, we impose
no-slip boundary conditions on the velocity fields i.e. $\bf v=0$ at $z=\pm W/2$ for all $x, y$. To proceed further, we assume that the dependence of the in-plane flow field
on the $z$ coordinate is Poiseuille for a given instantaneous in-plane
velocity at the mid-plane: We write
\begin{equation}
v_\alpha(x,y,z)=(z^2-W^2/4)\theta_\alpha (x,y),
\label{poise}
\end{equation}
which clearly satisfy the no-slip conditions at $z=\pm W/2$. Here, $\theta_\alpha (x,y)$ is a vector function containing the $(x,y)$ dependences of the $3d$ velocity fields $v_\alpha (x,y,z)$.
We assume $p_{y}(x,y,z)=p_{y}(x,y)$, i.e. $p_{y}$ has no $z$-dependence to the leading order. Using the boundary condition on $p_{z}$ we can write the simplest form of $p_{z}$ as
\begin{equation}
p_{z}(x,y,z)=\psi_{z}(x,y)\left(z^2-{W^2 \over 4}\right),
\end{equation}
where $\psi_z(x,y)$ determines the in-plane variation of $p_z(x,y,z)$. In our notations,
the fields $\theta_\alpha (x,y)$ and $\psi_z(x,y)$ are proportional to the mid-plane ($z=0$) values of the corresponding fields, $v_\alpha(x,y,z=0)$ and $p_z(x,y,z=0)$:
\begin{eqnarray}
v_\alpha (x,y,z=0)= -\theta_\alpha (x,y) W^2/4\equiv v_i (x,y),\\
p_z(x,y,z=0)=-\psi_(x,y)W^2/4,\\
p_y(x,y,z=0 )=p_y(x,y),
\end{eqnarray}
 since $p_y$ is assumed to be $z$-independent. Here $i=\alpha =x$ or $y$ and $v_i(x,y)$ is a $2d$ velocity field. With the choice of $z$-dependence of $p_y$ and $p_z$, the Franck free energy ${\mathcal F}_p$ in $3d$ becomes
\begin{equation}
{\mathcal F}_p=\frac{1}{2}\int dxdy \kappa[\frac{4\kappa}{3W}p_z^2 (x,y) + W(\partial_i p_y)^2].
\label{reducedfrank}
\end{equation}
Clearly, from the expression (\ref{reducedfrank}) above, $p_z$ is massive and does not survive in the long time limit, i.e., $p_z$ is not a slow variable. This is due to that fact $p_z$ is held fixed at the walls at $z=\pm W/2$. In contrast, $p_y$ is a massless field and its fluctuations are long lived; $p_y$ remains a slow variable in the problem. Further, Eq.~(\ref{reducedfrank}) allows us to define the $2d$ effective conjugate field $h_{\perp i}$ as
\beq
h_{\perp i}=-\frac{\delta {\mathcal F}_p}{\delta p_i}.
\eeq
Now we average the force balance equation Eq.~(\ref{genstokes}) over the thickness $W$ to obtain an equation as a function of $x$ and $y$ only and are
written in terms of $2$d vectors. After averaging, in the lubrication approximation \cite{lub} we get
\begin{equation}
 {1 \over W}\int_{-W/2}^{W/2}\eta_2\nabla^2v_{\alpha}\; dz=-{8\eta_2 \over W^2}v_i(x,y).
\end{equation}
After performing $z$-averaging over other terms, we find
for the $y$-th component of the velocity
\begin{eqnarray}
\partial_y \Pi&=&-{8\eta \over W^2}v_y-\xi'\partial_x(p_xp_y)\Delta\mu  + {\nu_1-1 \over 2W}\partial_x(p_xh_{\bot y}) \\
\Rightarrow v_y&=&-D\partial_y \Pi - \xi_{2} \partial_x(p_xp_y)\Delta\mu + \nu_{2}\partial_x(p_x h_{\bot y}),
\label{vy}
\end{eqnarray}
where constants $D={W^2 \over 8\eta}$, the inverse friction coefficient, $\xi_{2}=\xi'D$ and $\nu_{2}={\nu_1-1 \over 2W}D$. Similarly, for the $x$-component we get after averaging over $z$
\begin{eqnarray}
\partial_x \Pi&=&-{8\eta \over W^2}v_x-\xi'\partial_y(p_xp_y)\Delta\mu  + {\nu_1+1 \over 2W}\partial_y(p_xh_{\bot y}) \\
\Rightarrow v_x&=&-D\partial_x \Pi - \xi_{0} \partial_y(p_xp_y)\Delta\mu + \nu_{0}\partial_y(p_x h_{\bot y})
\label{vx}
\end{eqnarray}
where $v_x$ and $v_y$ are components of the $2d$ vector $v_i$; $\xi_{0}=\xi'D$ and $\nu_{0}={{\nu_1+1} \over 2W}D$.

Having obtained the effective $2d$ Eqs. of motion for $v_x$ and $v_y$ above we now proceed to obtain the same for  $\bf p$.
Linearising about the reference state chosen and neglecting the terms higher order in smallness Eq.~(\ref{ons3dp}) reduces to,
\begin{eqnarray}
{\partial p_y \over \partial t} = {h_y \over \gamma_1} - {\nu_1+1 \over 2}(\partial_y v_x)p_x - {\nu_1 - 1 \over 2}(\partial_x v_y)p_x +\xi_A\partial_x p_y.
\end{eqnarray}
Averaging over the thickness $W$, we get
\begin{eqnarray}
{\partial p_y \over \partial t}= {h_y \over \gamma_0} -{\nu_1+1 \over 3}(\partial_y v_x)p_x - {\nu_1-1 \over 3}(\partial_x v_y)p_x +\xi_A\partial_x p_y,
\label{py1}
\end{eqnarray}
where $\gamma_0=\gamma_1 W$ and $p_x=1$.

From Eq. (\ref{vy}) and Eq. (\ref{vx}) we get
\begin{eqnarray}
\partial_x v_y&=& -D\partial_x\partial_y\Pi - \xi_{2} \partial_x^2(p_xp_y)\Delta\mu + \nu_{2}\partial_x^2(p_xh_y), \label{dvx} \\
\partial_y v_x&=& -D\partial_x\partial_y\Pi - \xi_{0} \partial_y^2(p_xp_y)\Delta\mu + \nu_{0}\partial_y^2(p_xh_y).
\label{dvy}
\end{eqnarray}

Putting Eq. (\ref{dvx}) and Eq. (\ref{dvy}) in Eq. (\ref{py1}) we get
\begin{eqnarray}
{\partial p_y \over \partial t}&=& {h_{\bot y} \over \gamma_0} - {2p_xW\nu_{0}^2 \over 3D}\partial_y^2(p_x h_{\bot y}) - {2p_xW\nu_{2}^2 \over 3D}\partial_x^2(p_x h_{\bot y})  + {2p_x^2W\nu_{0} \over 3D}\xi_{0}\Delta\mu\partial_y^2(p_xp_y)\nonumber \\&+& {2p_x^2W\nu_{2} \over 3D}\xi_{2}\Delta\mu\partial_x^2(p_xp_y) + (\nu_{0}+\nu_{2}){2p_xW \over 3}\partial_x\partial_y \Pi +\xi_A\partial_x p_y.
\label{interIp}
\end{eqnarray}
Clearly when $\Delta\mu=0$, Eqs.~(\ref{vy},\ref{vx},\ref{interIp}) do not immediately conform to the (symmetric) structure as stipulated by the Onsager Reciprocity Theorem. In order to achieve that we exploit the rescaling freedom of the fields that comes due to the (multiplicative) arbitrariness in the definition of $2d\; p_y$.
Let us scale $p_y$ by $\lambda$ where $\lambda$ is any real number. Then $p_y\rightarrow \lambda p_y$ and $h_{\bot y} \rightarrow \lambda h_{\bot y}$. Scale factor $\lambda$ is to be chosen such that for $\Delta\mu=0$, a fully symmetric structure of the coupled equations for $v_x,\,v_y$ and $p_y$ follow.
The scaled equations are
\begin{eqnarray}
v_y&=& -D\partial_y \Pi - \lambda\xi_{2} \partial_x(p_xp_y\Delta\mu) + \lambda\nu_{2}\partial_x(p_xh_{\bot y}),\label{vyy} \\
v_x&=& -D\partial_x \Pi - \lambda\xi_{0} \partial_y(p_xp_y\Delta\mu) + \lambda\nu_{0}\partial_y(p_xh_{\bot y}),\label{vxx} \\
{\partial p_y \over \partial t}&=& {h_{\bot y} \over \gamma_0} - {2p_xW\nu_{0}^2 \over 3D}\partial_y^2(p_x h_{\bot y}) - {2p_xW\nu_{2}^2 \over 3D}\partial_x^2(p_x h_{\bot y})  \nonumber \\
&&+ {2p_x^2W\nu_{0} \over 3D}\xi_{0}\Delta\mu\partial_y^2(p_xp_y) + {2p_x^2W\nu_{2} \over 3D}\xi_{2}\Delta\mu\partial_x^2(p_xp_y)+ (\nu_{0}+\nu_{2}){2p_xW \over 3\lambda}\partial_x\partial_y \Pi. \label{pyy}
\end{eqnarray}
We choose $\lambda=({2W \over 3})^{1/2}$. Thus Eq.~(\ref{pyy}) becomes
\begin{eqnarray}
\dot{p_y}&=&{h_{\bot y} \over \gamma_0} - {2W \over 3D}(\nu_{0}^2\partial_y^2+\nu_{2}^2\partial_x^2)h_{\bot y} + (\nu_{0} + \nu_{2})({2W \over 3})^{1/2}\partial_x\partial_y\Pi \nonumber \\
&&+ {2W \over 3D}\Delta\mu(\nu_{0}\xi_{0}\partial_y^2 + \nu_{2}\xi_{2}\partial_x^2)p_y +\xi_A\partial_x p_y.
\label{pyy1}
\end{eqnarray}

Now we rescale the coefficients $({2W \over 3})^{1/2}(\nu_{2},\nu_{0},\xi_{2},\xi_{0})\rightarrow(\nu_{2},\nu_{0},\xi_{2},\xi_{0})$. Hence, Eqs.~(\ref{vyy},\ref{vxx}) and (\ref{pyy1}) become
\bea
v_x &=& DF_x + \nu_0 \partial_y h_{\perp y} + \xi_0 \Delta\mu \partial_yp_y, \label{vxIfinal} \\
v_y &=& DF_y + \nu_2\partial_x h_{\perp y} + \xi_2\Delta\mu \partial_xp_y, \label{vyIfinal} \\
\dot{p_y}&=& [{1 \over \gamma} - {\nu_0^2 \over D}\partial_y^2 - {\nu_2^2 \over D}\partial_x^2]h_{\perp y} -\nu_0\partial_y F_x -\nu_2\partial_x F_y \nonumber \\
&&  -{\nu_0 \over D}\xi_0\Delta\mu\partial_y^2p_y - {\nu_2 \over D}\xi_2\Delta\mu\partial_x^2p_y +\xi_A\partial_x p_y.
\label{pyIfinal}
\eea
It is evident that effective $2d$ Eqs.~(\ref{vxIfinal},\ref{vyIfinal}) and (\ref{pyIfinal}) have a symmetric structure for $\Delta\mu=0$, as expected. Further, they follow the same symmetries as discussed at the beginning of the present section. An independent direct derivation of these $2d$ equations, based on symmetry arguments, is provided in Sec.~\ref{appen1}.

\subsection{System II: Polar order normal to the plane without external forces}
\label{configII}

We take the reference unperturbed state as
${\bf p_0}$=$(0,0,1).$
Thus the ordering is normal to the surface of the plane. Therefore, one has in-plane rotational symmetry in the system. We will see below the small fluctuations around this state allows for a $2d$ descriptions with polar symmetry without any in-plane polar order (the macroscopic polar order lies normal to the plane) and without any external forces. As before, we set $v_z=0$. Further, in the fluctuating state $\langle p_x\rangle=0=\langle p_y\rangle$ and $\langle v_x\rangle =0=\langle v_y\rangle$ (since there is no external force); thus $p_i$ and $v_i$ are of the same order of magnitude.
%
%
\begin{figure}[h]
\includegraphics[width=9cm]{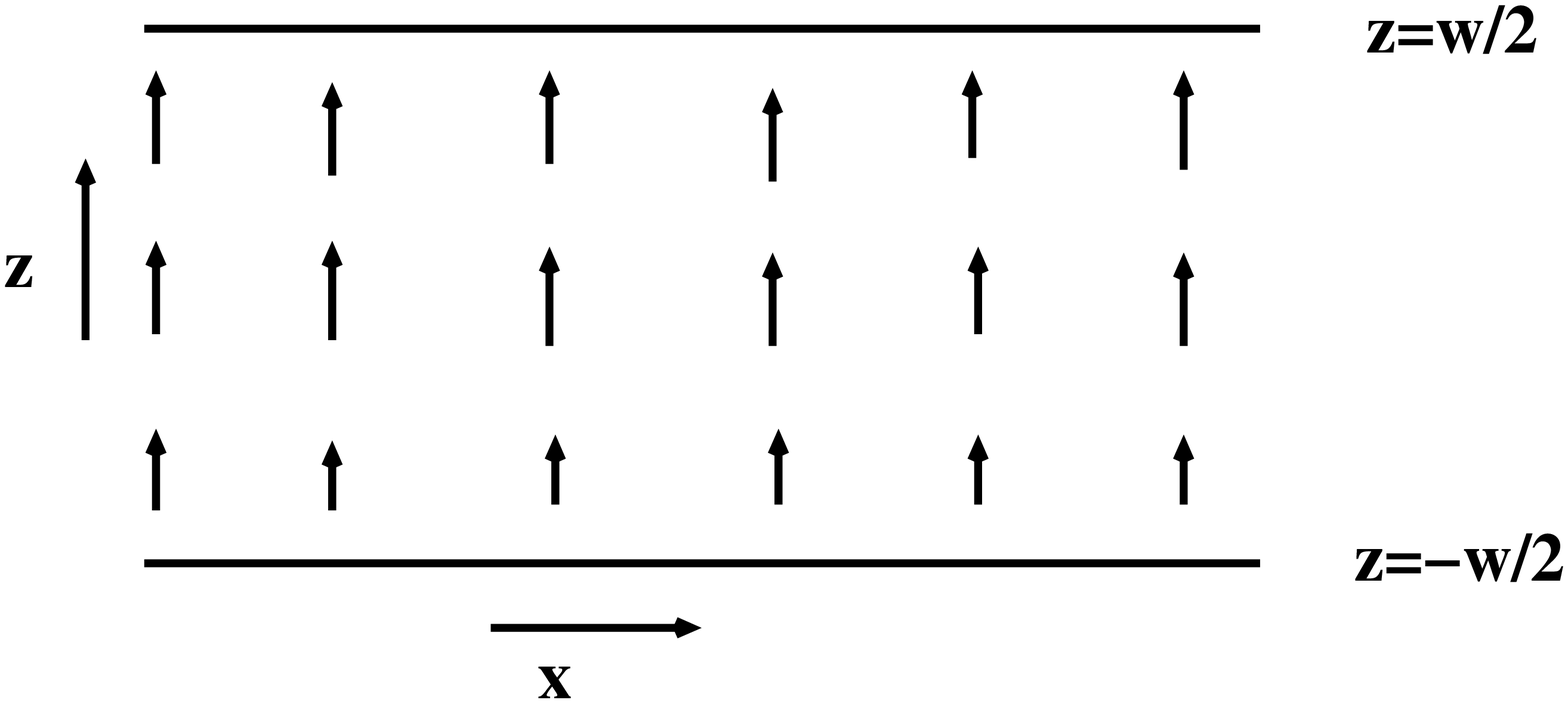}
\caption{A schematic diagram of the chosen initial reference state for System II. Arrows (parallel to the $z$-axis) indicate the direction of
$\bf p$ in the reference state.}
\end{figure}

Since $p_z=1$ in the unperturbed state and we are looking for
small fluctuations around it, fluctuations in $p_z$ are second order in smallness. Hence, the lowest order polarisation fluctuation $\delta\bf p$= $(p_x,p_y,0)$. Henceforth the dynamics of $p_z$
is ignored. We consider the dynamics of $p_x$ and $p_y$ as both are slow modes. We assume $p_x$ and $p_y$ to be independent of $z$ to the leading order and impose no anchoring conditions on them. Next we impose no-slip boundary condition on the velocity fields:
$v_\alpha(x,y,z=\pm W/2)=0$.
We impose the thin film approximation on the generalised Stokes Eq.~(\ref{genstokes}) and average over $z$. We find, as before,
$ {1 \over W}\int_{-W/2}^{W/2}\eta_2\nabla^2v_{\alpha}\; dz=-{8\eta_2 \over W^2}v_i(x,y)$.
In addition,
\begin{equation}
 {1 \over W}\int_{-W/2}^{W/2}u \; dz={1 \over 2}\int_{-W/2}^{W/2}\partial_\beta v_\beta dz\rightarrow\mbox{subleading}.
\end{equation}
All the other terms, upon $z$-averaging, turn out to be second order in
smallness. Therefore, new terms are to be added in the $3d$ onsager relations. The next order terms are polar terms and we have,
\begin{eqnarray}
\sigma&=&2\eta_1 u - \Delta \mu (\xi p^2+\bar{\xi}) + \bar{\nu}_1p_{\alpha}h_{\alpha} + \bar{\nu}_2\nabla\cdot{\bf h}
- \tilde{\xi}_2\Delta\mu\nabla\cdot{\bf p}, \\
\tilde{\sigma}_{\alpha\beta}&=&2\eta_2 \tilde{u}_{\alpha\beta}-\xi'\Delta\mu(p_{\alpha}p_{\beta}-{p^2 \over d}\delta_{\alpha\beta})
+ {\nu_1 \over 2}[p_{\alpha}h_{\beta}+p_{\beta}h_{\alpha}-{2 \over d}\delta_{\alpha\beta}p_{\gamma}h_{\gamma}] \nonumber \\
&&-{\xi_2'\Delta\mu \over 2}(\partial_{\alpha}p_{\beta}+\partial_{\beta}p_{\alpha}-{2 \over d}\nabla\cdot{\bf p}\delta_{\alpha\beta})
 + {\tilde{\nu}_2 \over 2}(\partial_{\alpha}h_{\beta}+\partial_{\beta}h_{\alpha}-{2 \over d}\nabla\cdot{\bf h}\delta_{\alpha\beta}),\\
{Dp_{\alpha} \over Dt}&=& {h_{\alpha} \over \gamma_1}+\lambda_1 p_{\alpha} \Delta\mu - \nu_1\tilde{u}_{\alpha\beta}p_{\beta}-\bar{\nu}_1 u p_{\alpha} +\bar{\nu}_2\partial_{\alpha}u
+ \tilde{\nu}_2\partial_{\beta}\tilde{u}_{\alpha\beta}.
\label{extons}
\end{eqnarray}

We henceforth set $\nu_1=\bar{\nu}_1=\xi=\bar{\xi}=\xi'=0$, since they do not contribute anything at the linear order. The modified force balance condition yields:
\begin{eqnarray}
\partial_{\alpha}\Pi&=&\eta_2\nabla^2v_{\alpha}+(\bar{\nu}_2-{\tilde{\nu}_2 \over d})\partial_{\alpha}\nabla\cdot{\bf h}+{\tilde{\nu}_2 \over 2}\partial_{\beta}(\partial_{\alpha}h_{\beta}+\partial_{\beta}h_{\alpha}) \nonumber \\
&&-({\xi}_2-{\xi_2' \over d})\Delta\mu\partial_{\alpha}\nabla\cdot{\bf p} -{\xi_2' \over 2}\Delta\mu\partial_{\beta}(\partial_{\alpha}p_{\beta}+\partial_{\beta}p_{\alpha}). \label{forcebal2}
\end{eqnarray}
Eq.~(\ref{forcebal2}) has a term ${\tilde{\nu}_2 \over 2}\partial_{\beta}(\partial_{\alpha}h_{\beta}+\partial_{\beta}h_{\alpha})$ which is equivalent to
${\tilde{\nu}_2 \over 2}\partial_j(\partial_ih_j+\partial_jh_i)$ as $p_z=1$ and $h_z=0$. Here $i,j=x, y$ only. In contrast $\dot{p_{\alpha}}$ has a term
${\tilde{\nu}_2 \over 2}\partial_{\beta}u_{\alpha\beta}$ which comes out to be ${\tilde{\nu}_2 \over 2}(\partial_z^2v_{\alpha}+2\partial_ju_{\alpha j})$.
Hence, to keep the force balance equation and $\dot{p_{\alpha}}$ on the same footing we modify the cross coupling term in the $\dot{p_{\alpha}}$ Eq. to $\tilde{\nu}_2\partial_ju_{\alpha j}, (j=x,y)$.
After the $z$ averaging Eq.~(\ref{forcebal2}), the equations of motion for the $2d$ velocity fields $v_i$ are obtained:
\begin{eqnarray}
v_i&=& -D\nabla_i\Pi + {\nu_{20} \over 2}\partial_j(\partial_jh_i+\partial_ih_j)+ \bar{\nu}_{20}\partial_i\nabla\cdot{\bf h}  - {\xi_{20} \over 2}\Delta\mu\partial_j(\partial_jp_i+\partial_ip_j)-\bar{\xi}_{20}\partial_i\partial_jp_j,
\label{3dcaseIIvfinal}
\end{eqnarray}
where $D={W^2 \over 8\eta_2}$, $\nu_{20}=\tilde{\nu}_2D$, $\bar{\nu}_{20}=(\bar{\nu}_2-{\tilde{\nu}_2 \over d})D$, $\bar{\xi}_{20}=(\tilde{\xi}_2-{\xi_2' \over d})D$ and $\xi_{20}={\xi}_2' D$.



After $z$ averaging the third of Eqs.~(\ref{extons}) resulting $2$d equation of motion of $p_i$ at the linear level is
\begin{equation}
\dot{p_i}={h_{\bot i} \over \gamma_0} + {2\tilde{\nu}_2 \over 3}\partial_ju_{ij} + {2 \over 3}(\bar{\nu}_2-\frac{\tilde{\nu}_2}{d})\partial_i u_i,
\label{3dcaseIIpinter}
\end{equation}
where $\gamma_0=\gamma_1W$.

In Eq.~(\ref{3dcaseIIpinter}) we substitute for $v_i$ by using Eq.~(\ref{3dcaseIIvfinal}) to obtain a constitutive relation for $p_i$:
\bea
\dot p_i &=&\frac{h_i}{\gamma_0}+\frac{\nu_{20}\tilde{\nu}_2}{6}\nabla^4_\bot h_{\bot i} + \left[\frac{\tilde{\nu}_2\nu_{20}}{2}+\frac{2\tilde{\nu}_2\overline\nu_{20}}{3} +\frac{2\nu_{20}}{3}(\overline \nu_2-\frac{\tilde{\nu}_2}{d}) + 2\frac{\overline\nu_{20}}{3}(\overline\nu_2-\frac{\tilde{\nu}_2}{d})\right]\nabla_i\nabla^2_\bot \nabla_\bot\cdot {\bf h}_\bot\nonumber \\& +& \frac{\tilde{\nu}_2D}{3}\nabla^2_\bot F_i +\left[\frac{\tilde{\nu}_2D}{3}+\frac{2D}{3}(\overline\nu_2 -\frac{\tilde{\nu}_2}{d})\right]\nabla_i \nabla_\bot\cdot {\bf F} - \frac{\tilde{\nu}_2}{6}\xi_{20}\Delta\mu\nabla^4_\bot p_i \nonumber \\
&&-\left[\frac{\tilde{\nu}_2\xi_{20}}{2} + {2\tilde{\nu}_2\bar{\xi}_{20} \over 3}\right]\Delta\mu \nabla^2_\bot \nabla_i
\nabla_\bot\cdot {\bf p}-{2 \over 3}(\bar{\nu}_2-{\tilde{\nu}_2 \over d})(\xi_{20}+\bar{\xi}_{20})\Delta\mu\nabla_i\nabla^2_\bot\nabla_\bot\cdot p.
\label{3dcaseIIpfinal}
\eea
Note that Eqs.~(\ref{3dcaseIIvfinal}) and (\ref{3dcaseIIpfinal}) do not really conform to the Onsager symmetry: In order to make the
system in agreement with the Onsager reciprocity theorem, we make the rescaling $p_i \rightarrow p_i\sqrt 2/\sqrt 3$. The final equations for $v_i$ and $p_i$ take the form:
\bea
v_i&=& -D\nabla_i\Pi + {\sqrt{2}\nu_{20} \over 2\sqrt{3}}\partial_j(\partial_jh_i+\partial_ih_j)+ {\sqrt{2}\bar{\nu}_{20} \over \sqrt{3}}\partial_i\nabla_\bot\cdot{\bf h} - {\sqrt{2}\xi_{20} \over 2\sqrt{3}}\Delta\mu\partial_j(\partial_jp_i+\partial_ip_j) - {\sqrt{2}\bar{\xi}_{20} \over \sqrt{3}}\nabla_i\nabla_\bot\cdot{\bf p},\nonumber \\
\dot p_i &=& \left[ {\delta_{ij} \over \gamma_0} + {\nu_{20}^2 \over 6D}(\nabla^4_\bot\delta_{ij}+ 3\nabla_i\nabla^2_\bot\nabla_j)+{2\nu_{20}\bar{\nu}_{20} \over 3D}\nabla_i\nabla^2_\bot\nabla_j + {2\nu_{20}(\nu_{20}+\bar{\nu}_{20}) \over 3D}\nabla_i\nabla^2_\bot\nabla_j\right]h_j \nonumber \\ &+& {\sqrt{2}\nu_{20} \over 2\sqrt{3}}\partial_j(\partial_jF_i+\partial_iF_j)
+ {\sqrt{2}\bar{\nu}_{20} \over \sqrt{3}}\nabla_i\nabla_jF_j - {\nu_{20}\xi_{20} \over 6D}\Delta\mu\left[\nabla^4_\bot p_i+ 3\nabla^2_\bot \nabla_i\nabla_\bot\cdot p\right] \nonumber \\ &-&{2\nu_{20}\bar{\xi}_{20} \over 3D}\Delta\mu\nabla_i\nabla^2_\bot\nabla_\bot\cdot p - {2\bar{\nu}_{20} \over 3D}(\xi_{20}+\bar{\xi}_{20})\Delta\mu\nabla_i\nabla^2_\bot\nabla_\bot\cdot {\bf p}.
\label{3dcaseIIfinal}
\eea

In Eq.~(\ref{3dcaseIIpinter}) one could have added another flow-orientation coupling term proportional to $\Delta\mu v_i$ which would have originated from a term of the form $\Delta\mu\partial_z^2 v_i$ in the $3d$ Eq. for $p_\alpha$. Such a term, being proportional to $\Delta\mu$, vanishes in the equilibrium limit and hence keeps the Onsager symmetry unchanged. When $\Delta\mu\neq 0$ this term generates a leading order active term $\Delta\mu \partial_i\Pi$ in Eq.~(\ref{3dcaseIIpfinal}), in addition to several subleading terms. Such terms however, being products of $\Delta\mu$ and $\partial_i\Pi$, beyond the scope of linear response regime.

By using the symmetry arguments given at the beginning of the section we can arrive at a similar set of equations directly, instead of averaging the $3d$ equations over the thickness. What results are equations identical to Eqs.~(\ref{3dcaseIIfinal}).
\bea
v_i &=& DF_i + \nu_2 \partial_i(\nabla_\bot\cdot {\bf h}_\perp) + \nu_3\nabla_\bot^2 h_i +\xi_2 \Delta\mu \nabla_\bot^2p_i+ \xi_3\Delta\mu\partial_i\nabla_\bot\cdot {\bf p},\label{case2vmu} \\
\dot{p_i}&=& \left[ {\delta_{ij} \over \gamma} + {(\nu_3 +\nu_2) \over D}\nu_2\nabla_\bot^2\partial_i\partial_j
+ {\nu_2{\nu_3} \over D}\nabla_\bot^2\partial_i\partial_j
 + {\nu_3^2 \over D}\nabla_\bot^4\delta_{ij} \right]h_{\perp j}  +\nu_2\partial_i(\nabla_\bot\cdot {\bf F})\nonumber \\
&& + \nu_3\nabla_\bot^2 F_i
+ {(\nu_2 +\nu_3) \over D}\xi_3\Delta\mu\nabla_\bot^2\partial_i(\nabla_\bot\cdot p) + {\nu_2 \over D}\xi_2\Delta\mu\nabla_\bot^2\partial_i(\nabla_\bot\cdot p)
+ {\nu_3\xi_2 \over D}\Delta\mu\nabla_\bot^4 p_i. \label{case2pmv}
\eea

Now comparing the above equation with the one obtained from Eqs.~(\ref{3dcaseIIfinal}) we get the following relationship among the various coefficients in the two different cases:
$
{1 \over \gamma} = {1 \over {\gamma_1 W}}, \,
\nu_3 = {\nu_{20} \over \sqrt{6}}, \,
\nu_2 = {\sqrt{2} \over \sqrt{3}}\left({\nu_{20} \over 2}+ \bar{\nu}_{20}\right), \,
\xi_2 = -{\xi_{20} \over \sqrt{6}}, \,
\xi_3 = -{\sqrt{2} \over \sqrt{3}}\left({\xi_{20} \over 2}+\bar{\xi}_{20}\right).
$

\subsection{System III: $2d$ planer polar order with external forces}
\label{configIII}

In this active gel film the $2d$ symmetry present in the sample is $y\rightarrow -y$. Since $p_x=1$, there is no invariance under $x\rightarrow -x$ and the system being polar it is also not invariant under ${\bf p}\rightarrow -{\bf p}$. Furthermore, $v_i$ is no longer a gradient due to the presence of external forces. The effective $2d$ equations of motion of this sample should be invariant under these symmetries. We write down an effective $2d$ description by averaging over the $3d$ sample.
In the above two cases we considered above the chosen reference states were uniform.
However, we now choose a non-uniform reference state. The reference state is given by,
\begin{eqnarray}
p_x&=&0,\; p_z=\pm1, \; p_y=0 \;\mbox{at} \; z=\pm W/2 \nonumber \\
p_x&=&1,\; p_z=0, \;p_y=0 \; \mbox{at} \; z=0 \nonumber \\
p_y&=&0 \; \mbox{everywhere in the bulk}
\end{eqnarray}
We parametrise the initial state $\bf p_0$ by
\begin{eqnarray}
p_x=\cos\theta(z), \; p_z=\sin\theta(z),\; p_y=0 \nonumber
\end{eqnarray}
\begin{figure}[h]
\includegraphics[width=9cm]{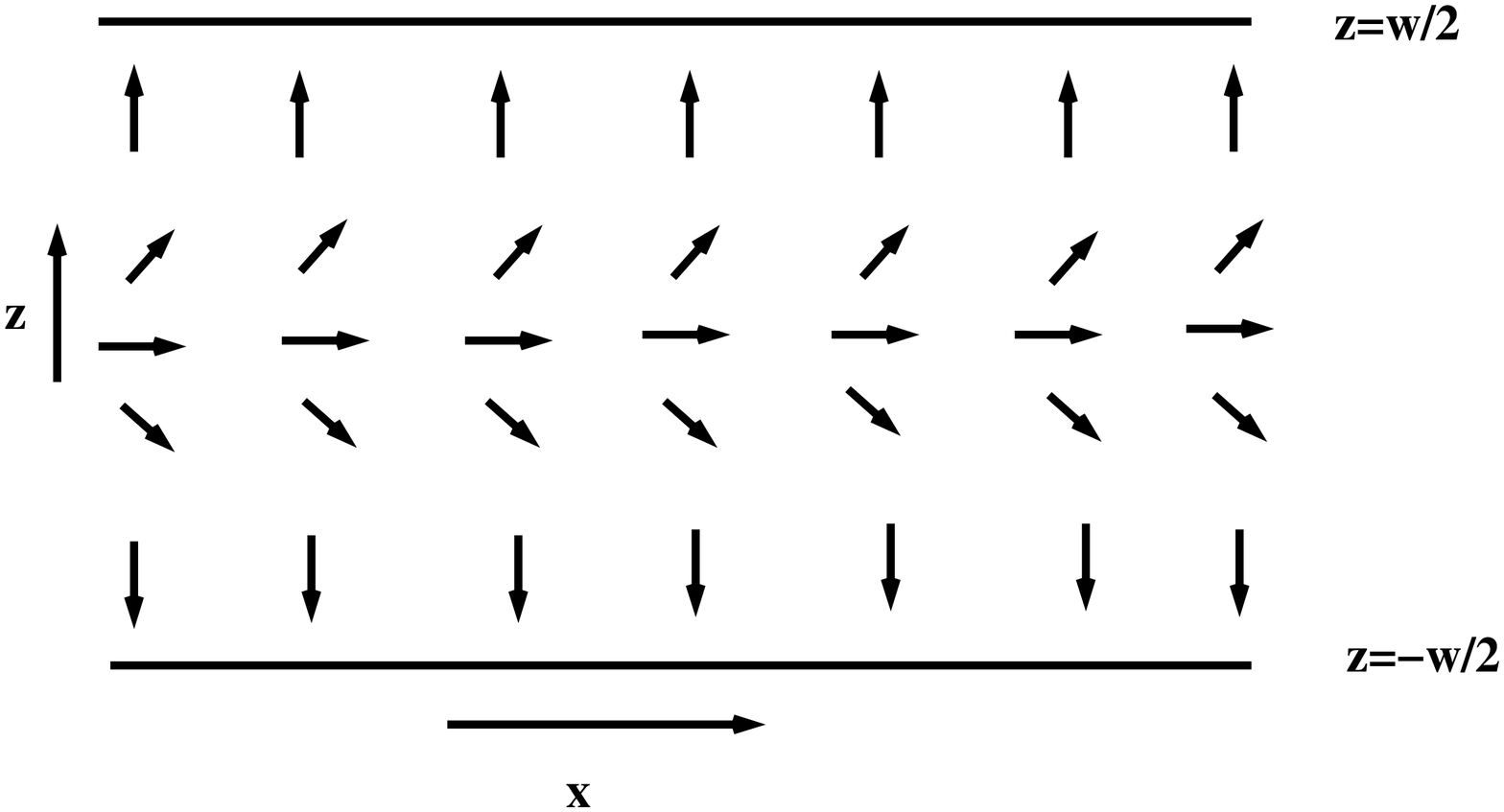}
\caption{A schematic diagram of the chosen initial reference state for System III. Arrows indicate the direction of
$\bf p$ in the reference state.}
\end{figure}
Before discussing fluctuations about this reference state in details, let us point out the difference between System III and Systems I and II. In the latter two, the transformation $p_x \rightarrow -p_x$ either makes $\bf p\rightarrow -p$ or $\bf p\rightarrow p$ in the reference state: In particular, this transformation for System I makes $\bf p\rightarrow -p$ (since $p_z=p_y=0$), and for System II makes $\bf p\rightarrow p$ (since $p_x=p_y=0$). Further, since microscopically the molecules are nematic, ${\mathcal F}_{3d} (p_x)={\mathcal F}_{3d}(-p_x)$. In contrast, the transformation $p_x\rightarrow -p_x$, when applied on System III, does not make $\bf p\rightarrow -p$. See Fig.~\ref{config3a} and Fig.~\ref{config3b} showing System III with $p_x\rightarrow -p_x$ and $\bf p\rightarrow -p$, respectively. Clearly, Fig.~\ref{config3a} and Fig.~\ref{config3b} are not the same. Nematic symmetry implies ${\mathcal F}_{3d} ({\bf p}) = {\mathcal F}_{3d} ({\bf -p})$. For Systems I and II, this immediately implies ${\mathcal F}_{3d} (p_x)={\mathcal F}_{3d} (-p_x)$. However, for System III, ${\mathcal F}_{3d}(p_x)\neq {\mathcal F}_{3d} (-p_x)$. Since effective $2d$ descriptions are constructed in terms of the mid-plane values of the $3d$ vectors, we expect ${\mathcal F}{p_x}\neq {\mathcal F}_{-p_x}$, indicating that the effective $2d$ descriptions for System III is expected to show up polar symmetry, which we confirm below.
\begin{figure}[h]
\includegraphics[width=9cm]{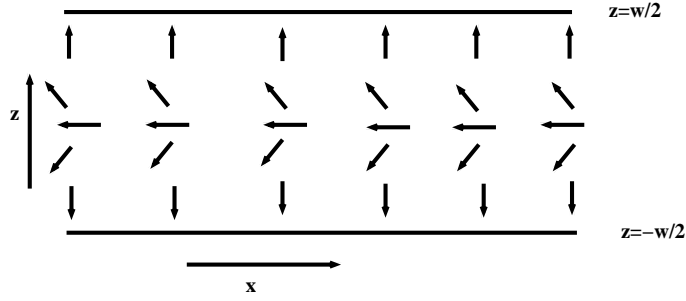}
\caption{System III with $p_x\rightarrow -p_x$.}
\label{config3a}
\end{figure}
\begin{figure}[h]
\includegraphics[width=9cm]{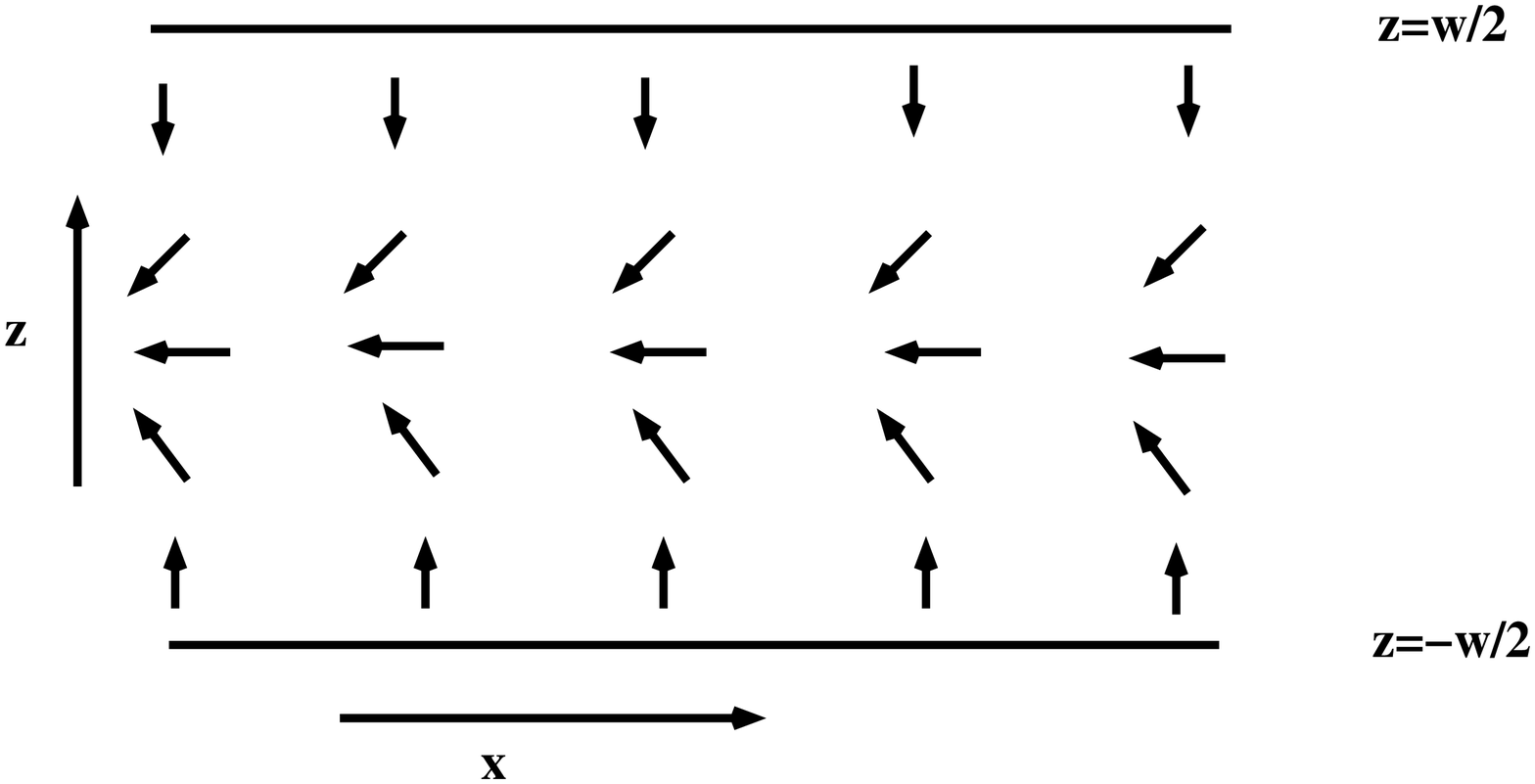}
\caption{System III with $\bf p\rightarrow -p$.}
\label{config3b}
\end{figure}
The Franck free energy in the equal Franck's constant limit becomes:
\begin{equation}
{\mathcal F}_p=-{\kappa \over 2}\int dxdy\int_{-W/2}^{W/2} \theta\nabla^2\theta dz=-\frac{\kappa}{2}\int dxdy \int_{-W/2}^{W/2}
\theta (z)\frac{d^2\theta}{dz^2}.
\end{equation}
for $p_x=\cos\theta(z), \; p_z=\sin\theta(z)$ and $p_y=0$.
The Euler Lagrange equation reads
\begin{equation}
 {d^2\theta \over dz^2}=0\;\;\Rightarrow\theta=Az+B.
\end{equation}
Imposing the boundary condition we obtain
 $\theta (z)={\pi z \over W}$, which thus defines the reference state.

We impose no slip-boundary conditions on the velocity fields. There is no restriction on $p_y$ (only $p_y$ is the slowest mode) at the boundaries.
Then $z$-averaging of the generalised Stokes Eq.~(\ref{genstokes}) together with the assumed $z$-dependence of the velocity field given by Eq.~(\ref{poise}) we obtain
\begin{eqnarray}
v_{ y}&=& -D\partial_y\Pi + \nu_{2}h_{\bot y} + \xi_{2}\Delta\mu p_{ y},\\
v_x&=&-D\partial_x\Pi - \nu_{0}\partial_yh_{\bot y} + \xi_{0}\Delta\mu\partial_yp_{ y},
\end{eqnarray}
where $\nu_{2}={\nu_1-1 \over W^2}D'$, $D={W^2 \over 8\eta}$ and $\xi_{2}={\xi'W \over 4\eta}$, $\nu_{0}={\nu_1+1 \over W\pi}D$ and $\xi_{0}={2\xi' \over \pi}D$.

Linearising Eq.~(\ref{ons3dp}) for the $y$-component about $\langle {\bf p}\rangle= \cos\theta(z){\bf e_x} + \sin\theta(z){\bf e_z}$ and averaging over the $z$ components we get
\beq
\partial_tp_{ y}= {h_{\bot y} \over \gamma_0} + 8{\nu_1-1 \over W\pi^2}v_{ y} + 8{\nu_1 + 1 \over \pi^3}\partial_yv_{ x},
\eeq
where $\gamma_0=\gamma_1W$.
Substituting for $v_{ x}$ and $v_{ y}$ in the above equation and rescaling $F_x=-{8W \over \pi^2}\partial_x\Pi$ and $F_y=-{8W \over \pi^2}\partial_y\Pi$, we get
\bea
v_x &=& D_xF_x + \nu_0 \partial_yh_{\perp y} + \xi_0\Delta\mu \partial_y p_y,\label{case3vx}\\
v_y &=& D_yF_y + \nu_2h_{\perp y} + \xi_2 \Delta\mu p_y,\label{case3vy}\\
\dot{p_y} &=& [{1 \over \gamma} + {\nu_2^2 \over D} - {\nu_0^2 \over D}\partial_y^2]h_{\perp y} + \nu_2F_y \nonumber \\
&& - \nu_0\partial_yF_x  - {\nu_0 \over D}\xi_0\Delta\mu\partial_y^2p_y.
\label{case3-3d}
\eea
where $\gamma = \gamma_1W ,
D_x=D_y = D$.
In Eqs.~(\ref{case3vx}) and (\ref{case3vy}) although we have $D_x=D_y$, in general $D_x\neq D_y$. See Sec.~\ref{appen2} for an alternative derivations, based on the $2d$ dymmetries, where such anisotropic coefficients appear naturally. Finally, one may generically add polar self-advecting terms $ \bf p\cdot \nabla p$ and $\bf p \nabla\cdot p$ in  the equation for $\dot{p_y}$ in Eq.~(\ref{case3-3d}). Upon linearisation, we find
\bea
\dot{p_y} &=& [{1 \over \gamma} + {\nu_2^2 \over D} - {\nu_0^2 \over D}\partial_y^2]h_{\perp y} + \nu_2F_y -\xi_A \partial_x p_y\nonumber \\
&& - \nu_0\partial_yF_x  - {\nu_0 \over D}\xi_0\Delta\mu\partial_y^2p_y,
\label{case3ons}
\eea
where $\xi_A$ is a coefficient of either sign.
Equations (\ref{case3vx} - \ref{case3ons}) formally constitute {\em effective} $2d$ description for System III.


\subsection{Dynamics of actin filaments in a motility assay}
\label{assay}

Motility assays are standard experimental set ups to measure properties of molecular motors such as myocins. In such an assay one observes in the light microscope the motion of isolated actin filaments being propelled by the myosin heads which are immobilised on a glass surface. Actin filaments are observed to slide over a layer of motor proteins (myosin) bound to the surface (glass plate). Actins bind with and unbind from myosin heads stochastically, resulting into velocity (or momentum) being imparted to the actin filaments. The actin velocity has a  mean and fluctuating parts. It has been observed that actin velocity increases with ATP concentrations \cite{hooft}. In  recent controlled experiments with many filaments at high densities, complemented by cellular automata based simulations \cite{erwin}, the authors demonstrated the spontaneous emergence of order and collective motion in the system. Further, they find large density fluctuations and propagating modes in the system. In this Sec. we discuss a simple coarse-grained description of the coupled dynamics of local actin velocity and polarisation degrees of freedom in a motility assay. We continue to use a one-fluid model, and thus do not distinguish between the solvent velocity and the velocity of the filaments (with the velocity field in our model equations corresponding to the centre-of-mass velocity in a two compoment system). This, although clearly an approximation, it suffices for our purpose to show the generic polar nature for our system and its connection (in the sense of same symmetries) with our {\em System III} discussed before. We assume that the actin filaments are preferentially oriented along the $x$-direction, i.e., $\langle {\bf p}\rangle =1$. Assuming incompressibility, we begin with the generalised Stokes Eq.~(\ref{genstokes}). However, in an actin motility assay, the velocity at one (or both) the walls is not zero, due to the forces imparted by the motors bound at the walls. Thus, instead of no-slip boundary conditions there is a finite slip velocity at the walls.
Without any loss of generality, we assume that the surface forces exist only at the wall at $z=W/2$. In the presence of the external forces, the force balance equation (\ref{forcebal}) is generalised to
\begin{equation}
\partial_\beta \sigma^{\rm tot}_{\alpha\beta} -\partial_\alpha \Pi = f^{\rm ext}_\alpha,
\label{genforcebal}
\end{equation}
where $f^{\rm ext}_\alpha$ is the total external force. These external surface forces, on symmetry grounds, should depend on either $h_\alpha$ or $p_\alpha$, and hence have two contributions of the form (i) $h_\alpha \delta (z-W/2)$ and $p_\alpha\Delta \mu \delta (z-W/2)$ in the hydrodynamic limit. The $\delta$-function ensures that the forces exist only on the top surface at $z=W/2$. The first contribution survives in equilibrium, while the second one is essentially a nonequilibrium contribution.

Together with these additional forces, the generalised Stokes equations for $v_x$ and $v_y$, that follows from the force balance equation (\ref{genforcebal}), for an actin motility assay become
\begin{eqnarray}
\eta\nabla^2 v_x &=& -\hat D\partial_x\Pi+ \xi' \partial_\gamma (p_x
 p_\gamma) - \frac{\nu_1}{2} \partial_\gamma (p_x h_\gamma + p_\gamma h_x) -\frac{1}{2}\partial_\beta (p_x h_\beta - p_\beta h_x) \nonumber \\ &+&
\tilde\nu_0h_x\delta(z-W/2) + \tilde\zeta_0 p_x\Delta\mu\delta(z-W/2),\\
\eta\nabla^2 v_y &=& -\hat D\partial_y\Pi+\xi' \partial_\gamma (p_y p_\gamma) - \frac{\nu_1}{2} \partial_\gamma (p_y h_\gamma + p_\gamma h_\beta) -\frac{1}{2}\partial_\beta (p_y h_\beta - p_\beta h_y) \nonumber \\ &+&
\tilde\nu_0h_y\delta(z-W/2) + \tilde\zeta_0 p_y\Delta\mu\delta(z-W/2),
\label{genstokes1}
\end{eqnarray}
where, coefficients $\tilde\nu_0$ and $\tilde\zeta_0$ couple the velocity fields with the surface force terms.
Averaging Eq.~(\ref{genstokes1}) over the thickness and in the lubrication approximation, we obtain for the $x$ and $y$-components of the $2d$ in-plane velocity field $v_i$
\bea
v_x &=& -\hat D\partial_x\Pi + \hat{\nu}_0h_{\bot x} + \hat\nu_x\partial_y (p_x h_{\perp y})+
 \hat{\xi}_0\Delta\mu p_{\bot x} + \hat\xi_x\partial_y (p_x p_y\Delta\mu), \label{assayveq1} \\
v_y &=& -\hat D\partial_y\Pi + \hat{\nu}_0h_{\bot y} + \nu_y\partial_x (p_x h_{\perp x}) +
\hat{\xi}_0\Delta\mu p_{\bot y} + \hat\xi_y\partial_x(p_xp_y\Delta\mu),
\label{assayveq}
\eea
with coefficients $\hat\xi_0,\,\hat\xi_x,\,\hat\xi_y,\hat\nu_0,\,\hat\nu_x$ and $\hat\nu_y$ which are related to the coefficients in Eqs.~(\ref{genstokes1}); $\hat D$ is an inverse friction coefficient.
To the leading order in gradients equations (\ref{assayveq1}-\ref{assayveq}) have exactly the same form as those of {\em Symmetry III} in Sec.~\ref{symm}. Clearly, velocities are
proportional to $\Delta\mu$ and the mean velocity $\langle v_x\rangle$ increases linearly with $\Delta\mu$, or, equivalently with the ATP concentration.
Since in this system $\langle\bar{p}\rangle = p_0\hat{e}_x=\hat{e}_x$ to the leading order, we ignore $h_{\bot x}$. Therefore,
\bea
v_x&=&-\hat{D}\partial_x\Pi + \hat\nu_x\partial_y (p_x h_{\perp y})+ \hat{\xi}_0\Delta\mu p_x + \hat\xi_x\partial_y (p_x p_y\Delta\mu),\nonumber \\
v_y&=& -\hat{D}\partial_y\Pi + \hat{\nu}_0h_{\bot y} + \hat{\xi}_0\Delta\mu p_{\bot y} +\hat\xi_y \partial_x(p_x p_y\Delta\mu).
\eea
Here, $\hat{\xi}_0\Delta\mu p_x$ gives a constant contribution to $v_x$. This term is responsible for the steady motion of the actin filaments as observed in actin motility assays. By comoving we can get rid of that. Therefore to the leading order,
\bea
v_x &=& \hat{D}F_x + \hat\nu_x\partial_y (p_x h_{\perp y}) + \hat\xi_x\partial_y (p_x p_y\Delta\mu),\nonumber \\
v_y &=& \hat{D}F_y + \hat{\nu}_0 h_{\bot y}  +\hat{\xi}_0 \Delta\mu p_y,
\label{vassay}
\eea
where we have used $F_i=-\nabla_i\Pi$. Equations~(\ref{vassay}) are identical to the constitutive relations set up for {\em Symmetry III} above.

The $3d$ bulk equation for the polar order parameter in an one component description is of the form \cite{joanny2}
\bea
\frac{\partial p_\alpha}{\partial t} + {\bf v}\cdot \nabla p_\alpha +\omega_{\alpha\beta} p_\beta +\xi_A {\bf p}\cdot \nabla p +\xi_B {\bf p} \nabla\cdot {\bf p}= \frac{h_\alpha}{\gamma_1} +\Lambda \nabla \Pi +\nu_1 u_{\alpha\beta}p_\beta,
\label{polarbulk}
\eea
where $\Pi_0$ is the pressure \cite{joanny2}
and $\Lambda$ is a coupling coefficient. To proceed further, we  linearise Eq.~(\ref{polarbulk}) about a given direction (say $x$-direction) for macroscopic orientation, we obtain an equation for $p_y$ of the form of Eq.~(\ref{case3ons}). This shows the close connections between our {\em System III} and the coarse-grained dynamics in an actin motility assay. More complete analysis of actin dynamics in a motility assay would require a two-fluid approach and proper boundary conditions on the free surface which we defer for a future work \cite{niladri2}.

\section{STOCHASTIC DYNAMICS AND TIME REVERSAL invariance}
\label{timerev}

The invariance of the statistical steady states under the time reversal operation characterises systems in thermal equilibrium. An
important consequence of this is the Fluctuation Dissipation Theorem (FDT) \cite{fdt} which connects correlation functions and
corresponding susceptibilities in equilibrium. For systems outside thermal equilibrium, there are no FDT. In order to calculate
correlation functions from the Eqs. of motion set up above, we add noises which represent the inherent stochasticity of the models
arising from the fast degrees of freedom which are ignored in the dynamical descriptions of the models. Stochastically driven coarse-grained
models have a long history in statistical mechanics, starting from equilibrium critical dynamics \cite{crit-dyn} and later on in nonequilibrium phenomena. Notable examples of the latter include surface growth phenomena \cite{kpz}, driven diffusive systems \cite{driven}, diffusion mediated reactions \cite{react} etc. In such stochastically driven models added noises are assumed to be zero-mean Gaussian distributed with specified variances. For systems in equilibrium, the
variances are fixed and are linked to the dissipative kinetic coefficients of the model. This arises as a consequence of the FDT. In systems of biological interests, the noises are of both thermal and non-thermal origin. In this article, we confine our discussions to the thermal noises
whose variances can be fixed by using the FDT for $\Delta\mu=0$.
We work out explicitly for {\em System I} below.
Introducing thermodynamic forces $F_x=-\partial_x\Pi$ and $F_y=-\partial_y\Pi$ the Onsager relations can be written in a
matrix notation as
\begin{eqnarray}
\left(
\begin{array}{c}
v_x \\
v_y \\
\dot{p_y}
\end{array}\right)=L_I \left(\begin{array}{c}
F_x \\
F_y \\
h_{\bot y}
\end{array}\right),\,\;L_{I}=\left(
\begin{array}{ccc}
D & 0 & i\nu_0q_y\\
0 & D & i\nu_2q_x\\
-i\nu_0q_y & -i\nu_2q_x & \frac{1}{\gamma} +\frac{\nu_2^2q_x^2}{D} +\frac{\nu_0^2q_y^2}{D}
\end{array}\right).
\end{eqnarray}
Hermiticity of $L_I$ follows from the fact that all the forces $F_x$, $F_y$ and $h_{\bot y}$ have the same property (even) under time reversal \cite{abmpi}. The time reversal property of the force can be found out from its definition. 
For a dissipative flux, its signature under time reversal is same as that of its corresponding force. Positivity of dissipation ensures that the dissipative coefficient relating a flux-force pair is strictly positive.
When $\Delta\mu=0$ the noise correlation matrix = $2k_BTL_I$ (see below). Such a choice ensures the FDT. When $\Delta\mu=0$ all the terms in the $v_x$ and $v_y$ equations are even under $t\rightarrow -t$. Since the active terms are proportional to $p_y$, they are odd under $t\rightarrow -t$. Thus $\Delta\mu$, a nonequilibrium drive, breaks the symmetry under the Onsager reciprocity principle. In order to show that our choice of the noise variance matrix $L$ indeed satisfies the FDT, we calculate the susceptibility and correlation matrices for the dynamical variables in the presence of external sources. For {\em System I}, the relevant dynamical variables are density $\rho$ and orientation field $p_y$. Their coupled dynamics may be written in a matrix notation as
\bea
\partial_t
\left(\begin{array}{c}
\rho\\p_y
\end{array}\right)=-\tilde L\chi_0^{-1}\left(\begin{array}{c}
\rho\\p_y
\end{array}\right) + \tilde L\left(\begin{array}{c}
\Pi_0\\h_{\perp y0}
\end{array}\right),
\eea
where $\Pi_0$ and $h_{y0}$ are externally imposed forces. Matrix $\chi_0$ is the static susceptibility matrix and in the Fourier space is given by
\bea
\chi_0^{-1}=\left(\begin{array}{cc}
\chi_\rho & i\alpha q_y\\
-i\alpha q_y & {\kappa q^2}.
\end{array}\right).
\label{susmat}
\eea
Matrix $\tilde L$ in the Fourier space is given by
\bea
\tilde L=\left[\begin{array}{cc}
Dq^2_\perp & - (\nu_0+\nu_2)q_xq_y\\
-(\nu_0 +\nu_2)q_xq_y & \frac{1}{\gamma} +\frac{\nu_0^2}{D}q_y^2 + \frac{\nu_2^2}{D}q_x^2
\end{array}\right].
\eea
The dynamic susceptibility matrix $\underline\chi$ is given by
$\underline\chi=\chi_{ij}(\omega)=[i\omega I + \tilde L\chi_0]^{-1}\tilde L$.
Here $I$ is the unit matrix. For a single-field model, the FDT yields a relation between the susceptibility $\tilde\chi (\omega)$ and correlation $\tilde C(\omega)$: $\chi''(\omega)=\frac{\beta}{2}\omega C(\omega),\,\beta=1/K_BT$ where $\chi''(\omega)$ is the imaginary part of $\chi(\omega)$. For a multi-component theory, such a relation generalises to a matrix equation and holds element-by-element:
$\chi_{ij}''(\omega)=\frac{\beta}{2}\omega C_{ij}(\omega)$,
where $i,j$ refer to different fields (here $\rho,\,p_y$), $\chi_{ij}''$ is the imaginary part of the susceptibility matrix element $\chi_{ij}(\omega)$. To calculate the correlation function matrix $\underline C\equiv C_{ij} (\omega)$, we add thermal noises $f_\rho \equiv -iq_x f_x -iq_y f_y$ and $f_p$, respectively, in the Eqs. for $\rho$ and $p_y$. Here, $f_x$ and $f_y$ are the thermal noises for $v_x$ and $v_y$. Noting that the variances of the noises $f_x,f_y,f_p$ are given by $2K_BTL_I$, we obtain the variances of the noises $f_\rho,f_p$ which is just $2K_BT\tilde L$.
Matrix $\underline C$ is then given by
\bea
\underline C(\omega)=\left(\begin{array}{c}
\rho \\ p_y
\end{array}\right)(\rho \;\; p_y)^* = [i\omega I + \tilde L\chi_0]^{-1}2K_BT\tilde L[-i\omega I +\chi_0\tilde L]^{-1}.
\eea
Further, from the definition of $C_{ij}(\omega)$ we have $C_{ij}(\omega)=C_{ji}(\omega)$ yielding $\chi_{ij}''(\omega) = -\chi_{ji}''(-\omega)$. We, therefore, conclude that $\chi_{ij}''$ is the {\em anti-hermitian} part of the full susceptibility matrix $\underline \chi$. One then obtains
\beq
\underline \chi'' = [i\omega I + \tilde L\chi_0]^{-1}\omega\tilde L[-i\omega I +\chi_0\tilde L]^{-1}=\frac{\beta}{2}\omega \underline C(\omega),
\eeq
establishing the FDT for the present case. This shows that our choice of the noise variance matrix is in agreement with the FDT.

\section{Linear stability analyses and correlation functions}
\label{eom}

In this Sec. we set up the Eqs. of motion for the relevant slow variables, starting from the $2d$ constitutive relations which we established above for the different cases. We separately consider the compressible and incompressible cases of the dynamics.

\subsection{Dynamics for {\em System I}}
\label{caseI}
We now set up the equations of motion of the slow variables for {\em System I}, for both nematic and polar orders. The slow variables are $p_y$ and mass density $\rho$ when the system is compressible; in the incompressible limit $\rho$, being a constant, drops out and $p_y$ remains the only slow variable in the problem. When $\Delta\mu=0$, the equations of motion are consistent with the Onsager reciprocity principle. The continuity equation for density $\rho$ reads
\begin{eqnarray}
{\partial \rho \over \partial t}=-\nabla\cdot(\rho{\bf v})=-[\partial_x(\rho v_x)+\partial_y(\rho v_y)].
\end{eqnarray}
Now linearising about the mean density $\langle\rho\rangle=\rho_0=1$ we get
\begin{eqnarray}
{\partial \rho \over \partial t}&=&D\nabla_\bot^2\Pi - (\nu_0+\nu_2)\partial_x\partial_yh_{\bot y} - \Delta\mu(\xi_2+\xi_0)\partial_x\partial_yp_y.
\label{eqdenI}
\end{eqnarray}

The free energy ${\mathcal F}_p$ for a polar sample, in the equal Franck's constant limit in $2d$ ($\kappa_1=\kappa_2=\kappa$), is given by
\begin{eqnarray}
{\mathcal F}_p=\int dxdy [-{\kappa \over 2}p_i\nabla_\bot^2p_i + \alpha{\bf p}\cdot\nabla_{\bot}\rho + {1 \over 2}\chi_\rho\rho^2 ]\equiv \int dxdy {\tilde f}_p.
\end{eqnarray}
Parameters $\alpha$ and $\chi_\rho$ are taken to be constants; $\alpha$ couples density fluctuations $\rho$ with polarisation $\bf p$, $\chi_\rho$ for a sample with
$\bf p=0$ is the inverse compressibility. For a nematic sample ${\mathcal F}_p$ must be invariant under $\bf p \rightarrow -p$ and hence $\alpha =0$ for a nematic sample; for a polar sample $\alpha\neq 0$.

From thermodynamic considerations we now use (linearising about a mean density $\rho_0$),
\begin{eqnarray}
\Pi=\rho{\delta {\tilde f}_p \over \delta\rho}-{\tilde f}_p=(\chi\rho - \alpha\nabla_{\bot}\cdot{\bf p}),\,
{\rm and}\,
h_{\bot y}=-{\delta {\mathcal F}_p \over \delta p_y}=(\kappa\nabla_{\bot}^2p_y + \alpha\nabla_{\bot}\rho)
\end{eqnarray}
Therefore, $\dot{p_y}$ in the Fourier space is written as,
\begin{eqnarray}
\dot{p_y}&=&[{-\kappa q^2 \over \gamma}-{\kappa \nu_0^2q_y^2q^2 \over D}- {\kappa \nu_2^2 q_x^2q^2 \over D} - i(\nu_0+\nu_2)\alpha q_xq_y^2  + {\nu_0 \over D}\xi_0\Delta\mu q_y^2 + {\nu_2 \over D}\xi_2\Delta\mu q_x^2 -i\xi_Aq_x]p_y\nonumber \\ &+&[{i\alpha q_y \over \gamma} + {i\alpha \nu_0^2 \over D}q_y^3 + {i\alpha\nu_2^2 \over D}q_x^2q_y - (\nu_0+\nu_2)q_xq_y\chi]\rho.
\label{pydynI}
\end{eqnarray}
Equations (\ref{eqdenI}) and (\ref{pydynI}) have been considered in Ref.~\cite{simha} in a discussion on active nematics.

The equations of motion for $\dot{p_y}$ and $\dot{\rho}$ can be cast in a matrix form whose eigenfrequency in the hydrodynamic limit (small q) is given by
\bea
\lambda(q_x,q_y)&=& {1 \over 2}[-Dq_\bot^2 - {\kappa q_\bot^2 \over \gamma} + {\nu_0\xi_0\Delta\mu q_y^2 \over D} + {\nu_2\xi_2\Delta\mu q_x^2 \over D} \pm \{ (-Dq_\bot^2 + {\kappa q_\bot^2 \over \gamma} - {\nu_0\xi_0\Delta\mu q_y^2 \over D} \nonumber \\
&& - {\nu_2\xi_2\Delta\mu q_x^2 \over D})^2 -4(\xi_0\nu_0 + \xi_0\nu_2 + \xi_2\nu_0 + \xi_2\nu_2)\Delta\mu q_x^2q_y^2\}^{1/2} ].
\label{geneigenI}
\eea
We can now compare eigenfrequencies (\ref{geneigenI}) with those in Ref.~\cite{simha}. Although our Eqs. for {\em System I} are invariant under the same set of symmetries as those in Ref.~\cite{simha}, there are some differences in details. For example, the {\em control parameter} $\alpha$ for creating instability in
Ref.~\cite{simha} appears in the cross coupling term in their density Eq.; in contrast, in our model $\Delta\mu$ plays similar role and appears in the orientation equation direction, in addition to the density equation where it appears as a cross-coupling coefficient. As a result, our model has more complicated behaviour (in regard to the presence of propagating modes, damping or instabilities) as a function of $\Delta\mu$, although the general scaling behaviour is identical to those in Ref.~\cite{simha}. One may further calculate
the two eigenvalues separately as functions of $(q_x,q_y=0)$ and $(q_x=0,q_y)$ and analyse their stability:
\begin{eqnarray}
\lambda(q_x,q_y=0)&=& -{\kappa q_x^2 \over \gamma}- {\kappa\nu_2^2q_x^4 \over D}+{\nu_2\xi_2\Delta\mu q_x^2 \over D}-i\xi_Aq_x, \; -Dq_x^2\chi,\label{eigIx}, \\
\lambda(q_x=0,q_y)&=& {1 \over 2D\gamma}[-D\kappa q_y^2-\gamma\kappa\nu_0^2q_y^4+\gamma\nu_0\xi_0\Delta\mu q_y^2-D^2\gamma\chi q_y^2]\nonumber \\
&&\pm \{(D\kappa q_y^2+ \kappa \gamma\nu_0^2 q_y^4 - \gamma\nu_0\xi_0\Delta\mu q_y^2 +D^2\gamma\chi q_y^2)^2 \nonumber \\
&&- 4D\gamma(-D^2\alpha^2q_y^4 - D\alpha^2\nu_0^2\gamma q_y^6 + D^2\kappa \chi q_y^4 + D\kappa\gamma\nu_0^2\chi q_y^6 \nonumber \\
&& - D\gamma\nu_0\xi_0\Delta\mu\chi q_y^4)\}^{1/2}.
\label{eigIy}
\end{eqnarray}
Equations (\ref{eigIx}) and (\ref{eigIy}) shows that depending upon the sign of the coefficients one or both the eigenvalue(s) may become unstable. For ${\nu_2\xi_2\Delta\mu \over D}>0$ and $|{\nu_2\xi_2\Delta\mu \over D}|>|{\kappa \over \gamma}|$ one of the eigenvalues $\lambda(q_x,q_y=0)$ changes sign and leads to low wavevector instability of the corresponding eigenvector. The other eigenvector remains stable. Therefore, physically only a specific linear combination of $p_y$ and $\rho$, determined by the unstable eigenvector, is unstable. However, there are no instabilities at ${\bf q}=0$ as both the eigenvalues vanish. Further, for large enough $\bf q$, both eigenvalues are negative. In case of equilibrium ($\Delta\mu=0$) both the eigenvalues are stable.

Having considered the linear instabilities, we now investigate the opposite situation when the system is linearly stable and admits nonequilibrium steady states. This happens when the condition of linear instability is not satisfied, i.e., when the active stress is either tensile ($\Delta\mu >0$ in our notation) or it is still contractile, but $|\Delta\mu| < |\Delta\mu_c|$.
The correlation function $S_q\equiv\langle \rho(q)\rho(-q)\rangle$ in the hydrodynamic limit can be calculated adding noise terms in the equation for $\dot{\rho}$ and $\dot{p_y}$. The Eqs.~(\ref{eqdenI}) and (\ref{pydynI}) take the form
\bea
\partial_t \rho &=& \chi D\nabla^2\rho -(\xi_0 +\xi_2)\Delta\mu\partial_x\partial_y p_y + \nabla_i\zeta_i, \nonumber \\
\partial_t p_y &=& \Gamma(\hat{{\bf q}})q^2p_y - (\nu_0+\nu_2)q_xq_y\rho + f_p.
\eea
where $\zeta_i$ is a Gaussian-distributed thermal noise for the velocity degrees of freedom with zero mean and a variance $\langle \zeta_i({\bf x},t)\zeta_j (0,0)\rangle = 2D\delta ({\bf x})\delta (t)$ and $\Gamma(\hat{{\bf q}})q^2=\left( -{\kappa q^2 \over \gamma} + {\nu_0\xi_0\Delta\mu \over D}q_y^2 + {\nu_2\xi_2\Delta\mu \over D}q_x^2\right)p_y$. We have considered the case when $\xi_A=0$. From these equations we obtain the equal-time correlation function (or structure factor)
$S_q\equiv\langle \rho(q)\rho(-q)\rangle = \int {d\omega \over 2\pi} \langle \rho(q,\omega)\rho(-q,-\omega)\rangle$.
Structure factor $S_q$ may show different behaviour depending upon whether or not propagating modes are present. For the choice of parameters when there are no propagating modes,
we find after a straight forward algebra
\bea
S_q\sim 1/q^2,
\eea
suggesting density fluctuations diverge as $L^2$ for a $2d$ system with a linear size $L$ \cite{simha}. In an equivalent equilibrium system this result would mean the compressibility $\chi$ diverging as $1/q^2$ in the $q\rightarrow 0$, or as $L^2$ in real space, where $L$ is the linear system size. Since $L^2\sim N$, where $N$ is the total number of particles, density fluctuations diverge as $N$ (see Ref.~\cite{simha,toner}). However, when  there are propagating modes (or when $\xi_A\neq 0$ and the system is polar),  such strong fluctuations get cut off.

\subsubsection{Incompressible limit}

\begin{figure}[h]
\includegraphics[width=7cm]{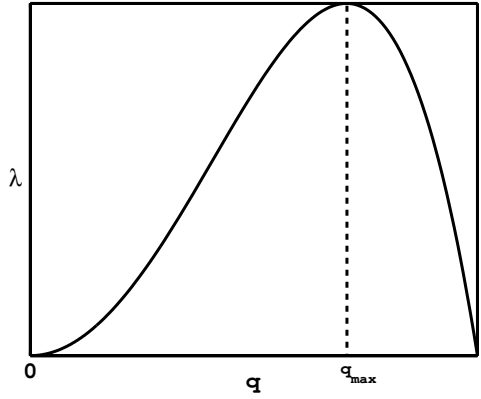}
\caption{A schematic plot of the real part of unstable eigenvalue $\lambda$ versus wavector $q$. The vertical broken line gives a measure of $q_{max}$.}
\label{fig1a}
\end{figure}

\begin{figure}[h]
\includegraphics[width=7cm]{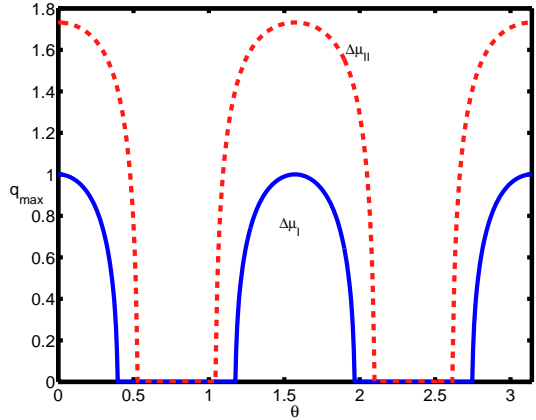}
\caption{A schematic plot of $q_{max}$ versus $\theta$ for some values of the parameters in {\em System I} for two values of $\Delta\mu: \Delta\mu_{II}>\Delta\mu_I$ (see text).}
\label{q1}
\end{figure}

In typical biologically relevant situations, flow velocities are of very small magnitudes (typically much smaller than the sound speed). As a result, the system behaves as an incompressible system and we set density $\rho=1$.
Incompressibility is enforced by $\nabla\cdot{\bf v}=\partial_xv_x+\partial_yv_y=0$. Using this we can express pressure  in terms of the other remaining fields in the problem and eliminate it from the dynamics of the problem. We have
\begin{eqnarray}
\Pi={1 \over D\nabla^2}[(\nu_0+\nu_2)\partial_x\partial_yh_{\bot y} + (\xi_0+\xi_2)\Delta\mu\partial_x\partial_yp_y].
\label{pressure1}
\end{eqnarray}
Substituting for $\Pi$ from equation (\ref{pressure1}), we
therefore obtain for ${p_y}$ in the Fourier space (in the polar coordinates $q_x=q_{\bot}\cos\theta$ and $q_y=q_{\bot}\sin\theta$)
\begin{eqnarray}
\dot{p_y}&=& -{\kappa q_{\bot}^2p_y \over \gamma} - {\kappa\over D}q^4(\nu_0 \sin^2\theta -\nu_2\cos^2\theta)^2 p_y
+ \frac{\Delta\mu}{D}q^2 (\nu_2\xi_2 \cos^4\theta +\nu_0\xi_0\sin^4\theta)p_y \nonumber \\
&& - \frac{\Delta\mu}{4D} q_\bot^2(\nu_0\xi_2 + \nu_2\xi_0)\sin^22\theta p_y -i\xi_A q\cos\theta p_y.
\label{polarpy1}
\end{eqnarray}
Eq.~(\ref{polarpy1}) is similar to the one derived in Ref.~\cite{sumithra} for fixed height and constant density.
Writing $p_y(q,t)\sim \exp[\lambda(q,\theta)t]$, we obtain for the eigenvalue $\lambda$
\begin{eqnarray}
\lambda(q,\theta)&=& -{\kappa q_{\bot}^2 \over \gamma} - {\kappa\over D}q^4(\nu_0 \sin^2\theta -\nu_2\cos^2\theta)^2
 + \frac{\Delta\mu}{D}q^2 (\nu_2\xi_2 \cos^4\theta +\nu_0\xi_0\sin^4\theta) \nonumber \\
&& - \frac{\Delta\mu}{4D} q_\bot^2(\nu_0\xi_2 + \nu_2\xi_0)\sin^22\theta  -i\xi_A q\cos\theta.
\label{lamcase1}
\end{eqnarray}
Clearly for $\Delta\mu=0$ there are no instabilities which is expected and, with $\xi_A=0$, equilibrium decay results are recovered, such that $\lambda(q,\theta,\Delta\mu=0) <0$. For non-zero $\Delta\mu$, two distinct cases arise: (i) When $\lambda (q,\theta) <0$, i.e., the system is stable, and (ii) when $\lambda(q,\theta)> 0$, i.e., the system is unstable. In the stable case, the role of $\Delta\mu$ is to enhance the decay rate of fluctuations. Further, there are propagating modes of speed $\xi_Aq \cos\theta$. Moreover, even when there are instabilities, the real part of $\lambda(q,\theta)$ initially rises from zero, reaches a maximum at $q=q_{max}$ and then decreases to eventually become negative. We find generally
\bea
q_{max}&=&(-\frac{D}{2\gamma (\nu_0\sin^2\theta -\nu_2\cos^2\theta)^2}  +\frac{\Delta\mu}{2\kappa (\nu_0\sin^2\theta -\nu_2 \cos^2\theta)^2}\nonumber \\&&\times\left[\nu_2\xi_2 \cos^4\theta + \nu_0\xi_0\sin^4\theta - \frac{1}{4}(\nu_0\xi_2 +\nu_2\xi_0)\sin^2 2\theta\right])^{1/2}.
\eea
A schematic plot of the real part of the unstable eigenvalue as a function of wavevector is shown in Fig.~\ref{fig1a}: It reaches a maximum and then comes down to zero to become negative eventually for high wavectors.
Since the instability peaks at wavenumber $q_{max}$, in the corresponding real space picture there will be a pattern periodic at a scale $\sim 1/q_{max}$. Since $q_{max}$ depends upon $\theta$, the generated pattern is clearly anisotropic. A schematic plot of $q_{max}$ versus $\theta$ is shown in Fig.~\ref{q1}. We clearly see (i) $q_{max}$ depends strongly on $\theta$, an illustration of the ensuing anisotropic pattern and (ii) for larger $\Delta\mu (\Delta\mu_{II}>\Delta\mu_I)$ $q_{max}$ is larger, implying that as $\Delta\mu$ rises patterns become denser in the real space.

\begin{figure}[h]
\includegraphics[width=7cm]{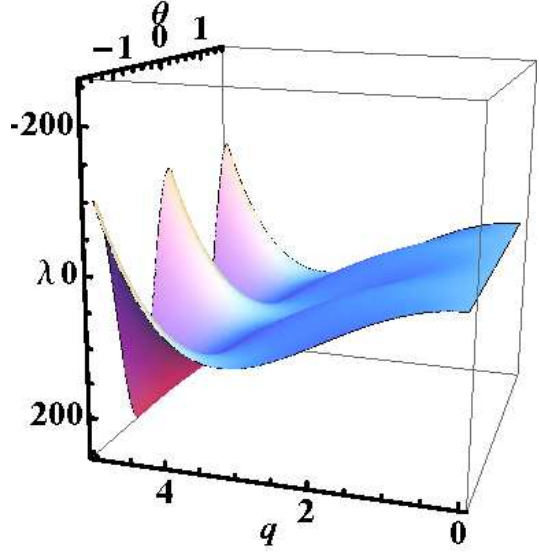}
\caption{(Colour online) A schematic $3d$ contour plot of the eigenvalue $\lambda$ as a function of wavevector $q$ and polar angle $\theta$ for some specific parameter values for {\em System I}. Both positive and negative values of $\lambda$ may be seen (see text).}
\label{eigen1}
\end{figure}

We analyse for a few different values of $\theta$. At $\theta=\pi/4$
\begin{eqnarray}
\dot{p_y}&=& -{\kappa q_{\bot}^2p_y \over \gamma} - {\kappa q^4\over 4D}(\nu_0-\nu_2)^2 p_y + \frac{\Delta\mu q^2}{4D}(\nu_2\xi_2  +\nu_0\xi_0 - \nu_0\xi_2 -\nu_2\xi_0) p_y  -\frac{i}{\sqrt 2}q\xi_A p_y.
\end{eqnarray}
In general however if
\begin{eqnarray}
|{k \over \gamma}q^2 - \frac{\Delta\mu q^2}{D}(\nu_2\xi_2 +\nu_0\xi_0) + \frac{\Delta\mu}{4D}(\nu_0\xi_2 +\nu_2\xi_0)q^2 + {\kappa q^4\over 4D}(\nu_0-\nu_2)^2|<0,
\label{inst1}
\end{eqnarray}
then there would be long wavelength instabilities, with $q_{max}^2=| [{k \over \gamma} - \frac{\Delta\mu }{D}(\nu_2\xi_2 +\nu_0\xi_0) + \frac{\Delta\mu}{4D}(\nu_0\xi_2 +\nu_2\xi_0)]/{\kappa \over 4D}(\nu_0-\nu_2)^2|$.
Next, consider $\theta=0$. Then,
\begin{eqnarray}
\dot{p_y}&=& -{\kappa q_\bot^2 \over \gamma}p_y -{\kappa \nu_2^2q_y^4 \over D}p_y + {\Delta\mu\nu_2\xi_2q^2 \over D}p_y -i\xi_Aq_\bot p_y
\end{eqnarray}
Thus there would be long wavelength instability if ${\xi_2\Delta\mu\nu_2 \over D} > 0$ and $|{\xi_2\Delta\mu\nu_2 \over D}| > {\kappa \over \gamma}$.
Threshold of instability is given by
${\xi_2\Delta\mu\nu_2 \over D} - {\kappa \over \gamma}=0.$
At $\theta = {\pi \over 2}$ also we get the same result as in $\theta=0$.
In general, since $\lambda$ depends explicitly on the polar angle $\theta$, it is entirely possible to have stable and unstable regions coexisting together for different values of $\theta$. For example, for angles determined by the inequality (with $\xi_0\Delta\mu >0$ and $\xi_2\Delta\mu>0$)
\bea
\frac{\kappa}{\gamma}+\frac{1}{4D}(\nu_0\xi_2 +\nu_2\xi_0)\Delta\mu \sin^2 2\theta < \frac{\Delta\mu}{D}(\nu_2\xi_2\cos^4\theta +\nu_0\xi_0\sin^4\theta),
\label{angle}
\eea
there will be instabilities, where as for angles outside this zone, it will be stable. On the other hand, for sign reversal of $\xi_0\Delta\mu$ and $\xi_2\Delta\mu$, $\theta$ satisfying Eq.~(\ref{angle}) corresponds to stable region and $\theta$ outside this domain corresponds to instability.
Thus, depending upon the parameter values, certain anisotropic (i.e., $\theta$-dependent) patterns will emerge in which $p_y$ will have large magnitudes for certain values of $\theta$, whereas $p_y$ will decay to zero for other values of $\theta$.  In each of these cases, replacement of the inequality sign by the equality sign in the relation (\ref{angle}) yields a critical activity $\Delta\mu_c$ above which the instability sets in. A contour plot of $\lambda$ versus $q$ and $\theta$ is shown in Fig.~\ref{eigen1} clearly depicting negative (stable) and positive (unstable) regions.

In terms of the relations between the different coefficients as shown in Sec.~\ref{config1}, we can rewrite the instability condition (\ref{inst1}), which shows that there are different critical thicknesses along different directions of the $XY$-plane. In particular, we find for $\theta=0$ the critical thickness $W_{c0}$ is given by
\beq
W_{c0}=\left[\frac{-24\eta\kappa}{\gamma_1(\nu_1-1)\xi'\Delta\mu}\right]^{1/2},
\eeq
and for $\theta=\pi/2$, the critical thickness $W_{c\pi/2}$ is
\beq
W_{c\pi/2}=\left[\frac{-24\eta\kappa}{\gamma_1(\nu_1+1)\xi'\Delta\mu}\right]^{1/2}.
\eeq
Since $W_{c0}\neq W_{c\pi/2}$, instabilities set in at different thickness in different regions of the $XY$-plane for a given $\Delta\mu$ [see Fig.~(\ref{eigen1})].

The correlation function in the incompressible case has to be $\langle p_y^2({\bf r},t)\rangle$ as $p_y$ is the only relevant flux in the system. We add a noise $\psi_p$ to Eq.~(\ref{polarpy1}) whose correlation is chosen so as to ensure FDT for $\Delta\mu=0$.
\bea
\langle \psi_p(q_{\bot}, \omega)\psi_p(-q_{\bot}, -\omega)\rangle &=& 2k_BT\left[{1 \over \gamma} +\frac{q_\bot^2(\nu_0 \sin^2\theta - \nu_2 \cos^2\theta)^2}{D}\right].
\eea
Correlation $\langle p_y^2({\bf r},t)\rangle$ reads
$\langle p_y^2({\bf r},t)\rangle = \int_{-\infty}^{\infty} {d\omega \over 2\pi}{d^2q_{\bot} \over (2\pi)^2}\langle |p_y({\bf q}_{\bot},\omega)|^2\rangle.$
Now from this, using the value of $\langle |\psi_p|^2\rangle$, we get
\bea
\langle p_y^2({\bf r},t)\rangle &=& k_BT\int {q_\bot dq_\bot \over 2\pi}{d\theta \over 2\pi}\left[{1 \over \gamma} +  \frac{q_\perp^2}{D}(\nu_0\sin^42\theta - \nu_2\cos^2\theta)^2 \right]\nonumber \\
&\times &[{\kappa q_{\bot}^2 \over \gamma} + \frac{\kappa q_\perp^4}{D}(\nu_0\sin^2\theta -\nu_2\cos^2\theta)^2 - \frac{\Delta\mu}{D}(\nu_2\xi_2 \cos^4\theta +\nu_0\xi_0\sin^4\theta)q_\bot^2 \nonumber \\ &+& \frac{\Delta\mu}{4D}(\nu_0\xi_2 +\nu_2\xi_0)q_\bot^2\sin^22\theta]^{-1}.
\eea
If $[{kq_\bot^2 \over \gamma} - \frac{\Delta\mu}{D}(\nu_2\xi_2 \cos^4\theta +\nu_0\xi_0\sin^4\theta)q_\bot^2  + \frac{\Delta\mu}{4D}(\nu_0\xi_2 +\nu_2\xi_0)q_\bot^2\sin^22\theta] > 0$ for all $\theta$, the system is stable and $\langle p_y^2({\bf r},t)\rangle$ has a logarithmic infrared divergence. In contrast when $[{kq_\bot^2 \over \gamma}- \frac{\Delta\mu}{D}(\nu_2\xi_2 \cos^4\theta +\nu_0\xi_0\sin^4\theta)q_\bot^2  + \frac{\Delta\mu}{4D}(\nu_0\xi_2 +\nu_2\xi_0)q_\bot^2\sin^22\theta ] = 0$, $\langle p_y^2({\bf r},t)\rangle$ has a quadratic infrared divergence.

In our formulation of the active gel problem the diffusion coefficients $D_{xx}$ and $D_{yy}$ for motions along the $x$- and $y$-directions can be calculated from the auto correlation functions $\langle v_i({\bf q},\omega) v_j ({\bf -q},-\omega)\rangle$ of the velocity field $\bf v$: In general $D_{ij}=\frac{1}{2}\int \frac{d^2q_{\perp}}{(2\pi)^d}_i({\bf q},\omega=0) v_j ({\bf -q},-\omega=0)\rangle$. The equations of motion of $v_x$ and $v_y$ in the incompressible limit are given by
\bea
v_x &=& - i(\nu_0 P_{xx}q_y\kappa q_{\bot}^2 + \nu_2P_{xy}q_x\kappa q_{\bot}^2 - \xi_0\Delta\mu P_{xx}q_y  - \xi_2\Delta\mu P_{xy}q_x)p_y + P_{xj}f_j, \nonumber \\
v_y &=& -i(\nu_2 P_{yy}q_x\kappa q_{\bot}^2 + \nu_0 P_{xy}q_y\kappa q_{\bot}^2 - \xi_2\Delta\mu P_{yy}q_x  - \xi_0\Delta\mu P_{xy}q_y)p_y + P_{yj}f_j,
\label{veqsymI}
\eea
where we have added thermal noise $f_i$ which is a zero-mean Gaussian noise with variances
\bea
\langle f_i(q_\perp,\omega)f_j(-q_\perp,-\omega)\rangle&=&2Dk_BT\delta_{ij},
\label{vncorr}
\eea
and $P_{ij}=(\delta_{ij}-\frac{q_i q_j}{q^2})$ is the transverse projection operator. Correlators $\langle v_i({\bf q},\omega) v_j ({\bf -q},-\omega)\rangle$ can be calculated from Eqs.~(\ref{veqsymI}) in a straightforward way. Evidently, off-diagonal elements $D_{xy}=D_{yx}$ are zero. Coefficients $D_{xx}$ and $D_{yy}$ have parts which depend on activity $\Delta\mu$ and have infra-red divergence. Ignoring the finite parts, we obtain
\bea
D_{xx}&=&{1 \over 2}\int {d^2q_{\bot} \over (2\pi)^2}\langle v_x(q_{\bot},\omega=0)v_x(-q_{\bot},\omega=0)\rangle \nonumber \\
&=& k_BT\int {d^2q_\bot \over (2\pi)^2}\left[\nu_0 P_{xx}q_yKq_{\bot}^2 + \nu_2P_{xy}q_xKq_{\bot}^2 + \xi_0\Delta\mu P_{xx}q_y + \xi_2\Delta\mu P_{xy}q_x\right]^2 \nonumber \\&\times&\left[{1 \over \gamma} + \frac{q_\perp^2}{D}(\nu_0\sin^2\theta -\nu_2\cos^2\theta)^2\right]|\Delta(\omega=0)|^{-2}.
\eea
where
\bea
\Delta(q_{\bot},\omega)&=& i\omega +{\kappa q_{\bot}^2 \over \gamma} + {\kappa \over D}(\nu_0q_y^2-\nu_2q_x^2)^2   - \frac{\Delta\mu}{Dq^2_\perp}(\nu_2\xi_2 q_x^4 +\nu_0\xi_0q_y^4) \nonumber \\
&& + \frac{\Delta\mu}{Dq^2_\perp}(\nu_0\xi_2 +\nu_2\xi_0)q_x^2q_y^2 + i\xi_A q_x.
\eea
We find that
if $\xi_A\neq 0$, then $D_{xx}$ has no infrared divergence. For $\xi_A=0$ there are contributions to $D_{xx}$ which are logarithmically infrared divergent. Further as $\Delta\mu\rightarrow \Delta\mu_c$, the critical $\Delta\mu$ for threshold of linear instability, $D_{xx}$ diverges as $(\Delta\mu_c -\Delta\mu)^{-2}$. In contrast in equilibrium, for a passive system, $D_{xx}$ is finite.   The diverging contribution in the active system stems from the explicit dependence of $v_x$ on $p_y$ and the fact that $p_y$-correlations are long-ranged in the plane. Moreover, using relations between the different coefficients as shown in Sec.~\ref{config1} we can draw further conclusions about the diffusion coefficient $D_{xx}$ (or, $D_{yy}$) of a tracer particle. The divergence of $D_{xx}$, now relates the critical thickness $W_c$ with the critical value of the activity $\Delta\mu_c$ as $W_c\sim 1/\sqrt \Delta\mu$. Thus as the $\Delta\mu$ increases, the corresponding critical thickness decreases (see Ref.~\cite{ab2} for more details). This allows one to experimentally test our results in which a thickness less than $L_c$ may be used and $D$ may be measured as $W\rightarrow W_{c-}$. Writing $W=W_c-\delta$ with $\delta/W_c \ll 1$, we obtain $D_{xx}\sim 1/\delta^2$ in the limit $\delta \rightarrow 0$, when the {\em active} contribution dominates. Diffusion coefficient $D_{yy}$ shows similar dependence on $\Delta\mu$ or $\delta$, although the numerical coefficient is different.

\subsection{Dynamics for System II: Incompressible limit}
\label{case2incom}

We begin with the constitutive relations given by Eqs.~(\ref{case2vmu}) and (\ref{case2pmv}). As in  System I (see Sec.~\ref{caseI} above), we use the Franck free energy to calculate ${\bf h}$. We work in the limit of equal Franck's constants. We have $h_{\perp i} = \kappa\nabla_\bot^2 p_i$. We are interested in the
incompressible limit of the dynamics and we eliminate pressure $\Pi$ using $\nabla_\perp\cdot{\bf v}=0$. Pressure may be expressed as
$\Pi= {1 \over D\nabla_\bot^2}[(\nu_2+\nu_3)\nabla_\bot^2(\nabla_\perp\cdot {\bf h}_\perp) + (\xi_2+\xi_3)\Delta\mu\nabla_\bot^2(\nabla_\perp\cdot {\bf p})]$.
Substituting for $\Pi$, Eq.~(\ref{case2pmv}) becomes
\begin{eqnarray}
\dot{p_i} &=& \left[\frac{\delta_{ij}}{\gamma} +\frac{\nu_3^2}{D}\nabla_\bot^4 P_{ij}\right]h_j +\frac{\nu_3\xi_2}{D}\Delta\mu P_{ij}\nabla^4 p_j+ g_i,
\label{case2dyn}
\end{eqnarray}
where $P_{ij}=\delta_{ij}-\frac{\partial_i\partial_j}{\nabla^2}$ is the transverse projection operator which appears due to the divergence-free condition of the velocity fields. We have added a thermal noise term $g_i$ which is a zero-mean Gaussian distributed noise with a variance given by
\beq
\langle g_i ({\bf r},t)g_j ({\bf 0},0)\rangle = 2K_BT\left[\frac{\delta_{ij}}{\gamma} +\frac{\nu_3^2}{D}\nabla_\bot^4 P_{ij}\right]\delta({\bf r})\delta (t),
\eeq
where $\delta ({\bf r})$ is the $2d$ Dirac $\delta$-function. This choice for the variances of $\psi_i$ ensures that the FDT is held true for $\Delta\mu=0$. We continue to use the same thermal noises for non-zero but small $\Delta\mu$.

\begin{figure}[htb!]
\includegraphics[height=6cm]{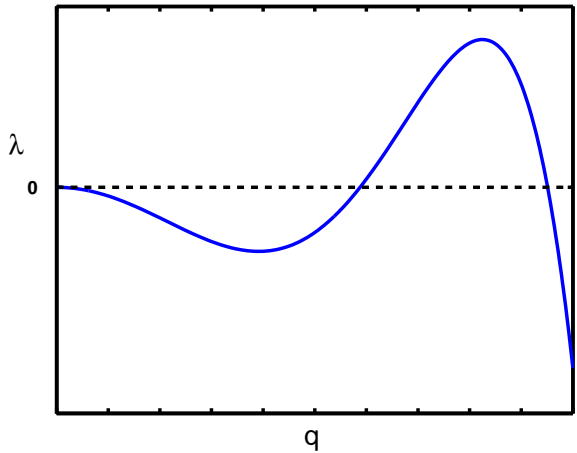}
\caption{(Colour online) A schematic plot of the eigenvalue $\lambda$ of the unstable mode versus wavevector $q$ for System II.}
\label{case2fig}
\end{figure}

In order to study linear instabilities, we ignore the noise and separate the individual components of $\dot{p_i}$ in fourier space and defining $p_i({\bf q},t)\sim p_i (t=0)\exp (\lambda t)$, we find 
\begin{equation}
\lambda=-{kq_\bot^2 \over \gamma} \mbox{,} -{kq_\bot^2 \over \gamma}-{\kappa\nu_3^2 \over D}q_\bot^6 + {\nu_3\xi_2\Delta\mu \over D}q_\bot^4.
\end{equation}
The eigenvalues $\lambda$ have no angle dependence. Therefore, the $2d$ rotational symmetry is maintained. Both the eigenvalues are stable at the longest wavelengths (smallest $q$). Further one of the eigenvalues is independent of $\Delta\mu$ and negative at all values of $q$, and hence, stable for all wavenumbers. The other one depends on $\Delta\mu$ and may change sign
(thus leading to instability) when $\nu_3\xi_2\Delta\mu<0$ for intermediate range wavenumbers given by $q_\bot^2{|\nu_3\xi_2\Delta\mu| / D}>{k \over \gamma}$ and the crossover is determined by $q_\bot^2{|\nu_3\xi_2\Delta\mu| / D}={k \over \gamma}$ yielding a crossover wavevector $q_{c1}$. For high enough wavenumbers (dominated by the $q^6$ term) the system
is again stable and asymptotically matches with the equilibrium results. The crossover to stability determined by the $q^6$ term is determined by the condition $q_\bot^2 \kappa\nu_3^2/D=|\nu_3\xi_2\Delta\mu / D|$ defining another crossover scale $q_{c2}$. Unlike {\em System I}, the eigenvalues have no angle dependence, and as a result, in the unstable case the growth rate is same in all directions in the $2d$ plane. This is a consequence of the $2d$ rotational symmetry. A schematic plot of the unstable eigenvalue $\lambda$ versus wavevector $q$ is shown in Fig.~\ref{case2fig}. Further, it is evident from the condition of the instability that for sufficiently small $\Delta\mu$ the intermediate band of unstable wavenumbers may vanish. The specific condition of threshold of instability is given by $q_{c1}=q_{c2}$.


In the linearly stable case,
the correlation functions of $p_x$ and $p_y$ in the incompressible limit may be calculated exactly in a straight forward way starting from stochastically driven Eqs. of motion of $p_x$ and $p_y$ (\ref{case2dyn}).
However, we do not present the detailed but algebraically rather intensive calculation here. Instead we provide arguments about the correlations in the scaling level. Fields $p_x$ and $p_y$ are coupled and they may be expressed in terms of eigenmodes whose time evolutions are independent of each other and are governed by the two eigenvalues of the matrix $M$ calculated above. As we find there, one of the eigenvalues is always stable and independent of $\Delta\mu$ and the other one is $\Delta\mu$-dependent and may be unstable depending upon the the value and sign of $\Delta\mu$. Although one of the eigenvalues is stable, since $p_x$ and $p_y$ are linear functions of the eigenmodes, both of them will be affected by the instability, if any, of the $\Delta\mu$-dependent eigenvalue. Linear relations between $p_x,\,p_y$ and the eigenmodes ensure that the nature of divergence is same as that of the correlation of the unstable eigenmode. In particular there are no instabilities at the longest wavelength. Since the eigenvalue which depends upon $\Delta\mu$ may change sign
(thus leading to instability) for $\nu_0\xi_0\Delta\mu<0$ for intermediate range wavenumbers given by $q^2{|\nu_0\xi_0\Delta\mu| \over D}>{k \over \gamma}$, the correlation function for the corresponding eigenmode and hence the correlation functions of both $p_x$ and $p_y$, will show divergence when the above condition is satisfied. In the compressible limit, the velocity fields $v_x$ and $v_y$ depend linearly on $p_x$ and $p_y$; hence velocity auto-correlators also will show divergences for wavevectors in the intermediate range satisfying the above mentioned conditions. This then means that the diffusion coefficient of a tagged particle in the sample, being proportional to spatial integrals over appropriate velocity correlation functions, will have divergences once the above instability conditions are met.

\subsection{Dynamics for System III}
\label{dynIII}

So far in the above we have set up the Eqs. of motion and analysed their instabilities when there are no external forces and the Eqs. for the velocity fields have a conservation law form. We now consider the dynamics for System III, which is the case when there are external forces, and, as a result, $\bf v$ no longer has a conservation law form. This will have important consequences on the ensuing dynamics as we will see below. We begin with the constitutive relations (\ref{case3vx}), (\ref{case3vy}) and (\ref{case3ons}). We consider the compressible limit in which case the conservation equation is given by
\bea
\dot{\rho}&=& -\nabla_\bot\cdot v = D_x\partial_x^2\Pi +D_y\partial_y^2\Pi - \nu_0\partial_x\partial_yh_{\bot y} - \nu_2\partial_yh_{\bot y} -\xi_0\Delta\mu\partial_x\partial_yp_y - \xi_2\Delta\mu\partial_yp_y.
\eea
Using the thermodynamic relations $\Pi=(\chi\rho-\alpha\nabla_{\bot}\cdot {\bf p})$ and $h_{\bot y}=(\kappa\nabla_\bot^2p_y+\alpha\partial_y\rho)$ as in {\em System I} we get
\bea
\dot{\rho} &=& -(D_xq_x^2+D_yq_y^2)\chi\rho +i\nu_0\alpha q_y^2q_x\rho+\nu_2\alpha q_y^2\rho +(i\alpha D_xq_x^2q_y+i\alpha D_yq_y^3 -\kappa\nu_0q_x^3q_y \nonumber \\
&& -\kappa\nu_0q_xq_y^3 + i\kappa\nu_2q_yq^2 + \xi_0\Delta\mu q_xq_y - i\xi_2\Delta\mu q_y)p_y - iq_xf_x - iq_yf_y.
\label{rho3}
\eea
where we have added zero-mean Gaussian distributed thermal noises $\zeta_i$ with a variance
$\langle \zeta_i(q,\omega)\zeta_j(-q,-\omega)\rangle = 2D\delta_{ij}$
and
\bea
\dot{p_y}&=& \left(-{\kappa q^2 \over \gamma} - {\kappa\nu_2^2q^2 \over D_y} - {\kappa\nu_0^2q_y^2q^2 \over D_x} + {\nu_0\xi_0\Delta\mu q_y^2 \over D_x} - \alpha\nu_2q_y^2 + i\alpha\nu_0q_xq_y^2 + i\xi_Aq_x \right)p_y + (i{\alpha q_y \over \gamma} \nonumber \\
&& + i{\alpha\nu_2^2 q_y \over D_y} + i{\alpha\nu_0^2q_y^3 \over D_x} -i\nu_2q_y\chi - \nu_0q_xq_y)\rho +f_p,
\label{py3}
\eea
where we have added a zero-mean Gaussian thermal noise $f_p$ with a variance
\bea
\langle f_p(q,\omega)f_p(-q,-\omega)\rangle = 2\left( {1 \over \gamma} + {\nu_2^2 \over D_y}\right).
\eea
Equations~(\ref{rho3}) and (\ref{py3}) have the same symmetry (polar) as those in Ref.~\cite{toner}. However, Eqs.~(\ref{rho3}) and (\ref{py3}) differ with the corresponding equations in Ref.~\cite{toner} in details.
Assuming $p_y(t),\,\rho(t)\sim \exp(\lambda t)$,
$\lambda$ may be calculated easily in a straight forward way.
From the expressions of the eigenfrequencies (we do not show the explicit forms here)  we find that there are dissipative (or instabilities) as well as underdamped propagating modes just like in {\em System I} depending upon the sign of the coefficients. In the large wavevector limit, both the eigenvalues are stable. {\em System III} is, therefore, characterised by the presence of generic polar order in the ordered state, large density and polarisation fluctuations together with moving instabilities. In this context we refer to Ref.~\cite{erwin}, where the authors experimentally and numerically studied the dynamics of actin filaments in a motility assay. For high enough density they found a polar ordered state with larger density fluctuations coupled with propagating waves. We believe our formulation here is a promising starting point for physical understanding of actin dynamics in a motility assay. Effects of nonlinear terms are likely to be important in understanding the density-dependent phase transition discussed in Ref.~\cite{erwin}.

When the system is linearly stable, as in Sec.~\ref{caseI}, we now examine the properties of relevant correlation functions in the ensuing nonequilibrium steady state. Again as in Sec.~\ref{caseI}, such situations arise when conditions for linear instability are not met.
For a compressible system, the density auto correlator shows novel behaviour, similar to but richer than for the case of {\em System I}.
We obtain the equal-time density auto correlation function $S(q,t)=
\int_{-\infty}^{\infty} \frac{d\omega}{2\pi}S ({\bf q},\omega)\equiv \langle \rho(q_\bot,\omega)\rho(-q_\bot,-\omega)\rangle$ as
 $S_{\bf q}\equiv\langle \rho ({\bf q},t)\rho (-{\bf q},t)\rangle = \int_{-\infty}^{\infty} S({\bf q},\omega) \sim 1/q^2$. The noticeable feature is that density fluctuations are enormous as $q\rightarrow 0$, diverging as $1/q^2$. This would be equivalent in an equilibrium system to having a compressibility which diverges as $1/q^2$. In real space terms, this would mean compressibility $\chi$ diverges as $L^2$ where $L$ is the linear dimension of the system. As we discussed before,irrespective of the details of a system or any connection to a response function, the rms number fluctuations $\sqrt {\langle\delta N^2\rangle}$ in an area $A$ (we are considering a $2d$ system) scales as $\sqrt {S(q\rightarrow 0)A}$. Thus for our system {\em System III} we have
\beq
\sqrt {\langle\delta N^2\rangle}\propto \sqrt {L^2 A}\propto L^2 \propto N,
\eeq
where, we have used $N\propto A\propto L^2$. Thus we find giant number fluctuations in the system \cite{toner}.

\subsubsection{Incompressible limit}

\begin{figure}[htb]
\includegraphics[height=5cm]{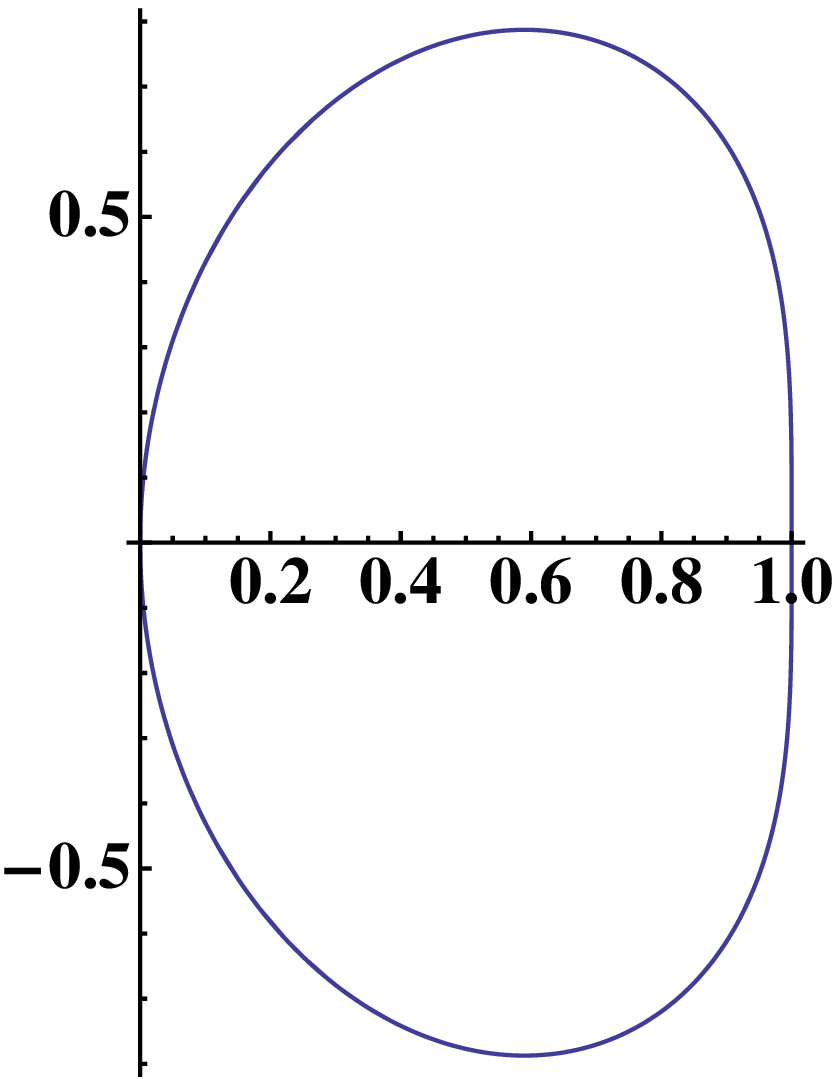}\\
\includegraphics[height=4cm]{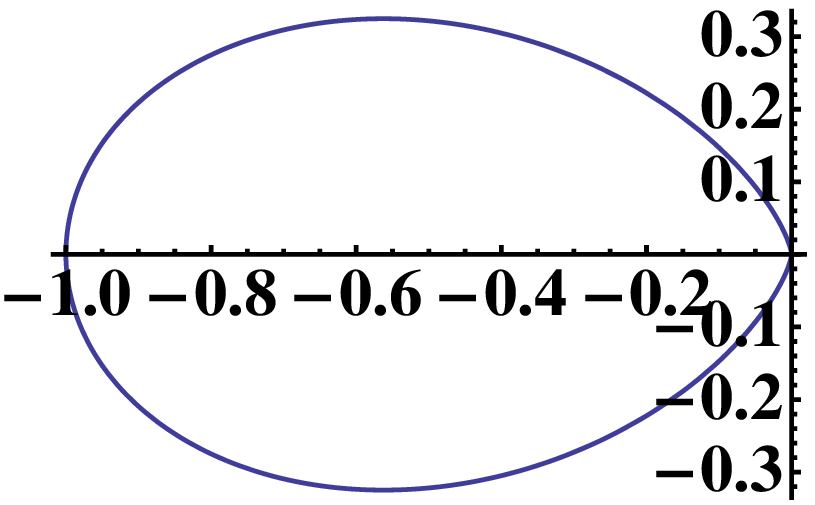}
\caption{A schematic polar plot of the wavespeed $C(q,\theta)$ as a function of $\theta$ for a given $q$. The anisotropic nature of the wavespeed is visible.}
\label{wave}
\end{figure}

\begin{figure}[h]
\includegraphics[width=7cm]{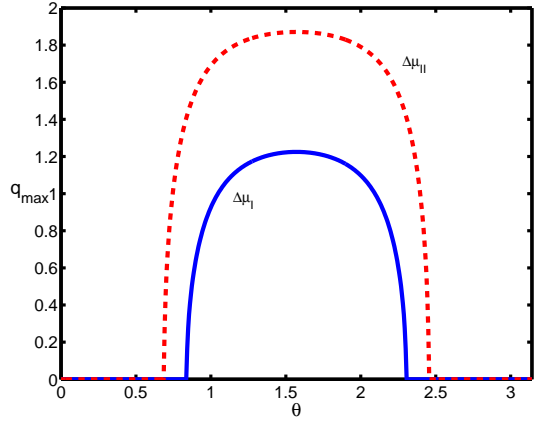}
\caption{(Colour online) A schematic plot of $q_{max}$ versus $\theta$ for some values of the parameters in {\em System III} for two values of $\Delta\mu: \Delta\mu_{II}>\Delta\mu_I$ (see text).}
\label{q3}
\end{figure}

\begin{figure}[htb!]
\includegraphics[width=7cm]{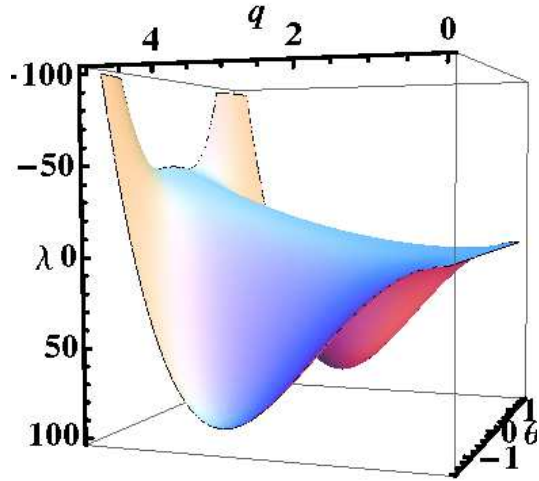}
\caption{(Colour online) A schematic $3d$ contour plot of the eigenvalue $\lambda$ as a function of wavevector $q$ and polar angle $\theta$ for some specific parameter values. Both positive and negative values of $\lambda$ are visible (see text).}
\label{eigen3}
\end{figure}

In this limit we take $\nabla\cdot {\bf v}=0$. Here the only slow variable is $p_y$ whose Eq. of motion we set up below. The pressure can be eliminated by using the incompressibility condition and we get
$\Pi = {1 \over D_xq_x^2 + D_yq_y^2}[\xi_0\Delta\mu q_xq_yp_y - i\xi_2\Delta\mu q_yp_y - \kappa\nu_0 q_xq_yq^2p_y + i\nu_2\kappa q_yq^2p_y]$.
Using this value of $\Pi$ we write Eq.~(\ref{case3ons}) as
\bea
\dot{p_y} &=& -{\kappa q_\bot^2 p_y \over \gamma} - {\kappa\nu_2^2q_\bot^2D_x\cos^2\theta - \kappa\nu_0^2q_\bot^4D_y\sin^4\theta\over D_y(D_x\cos^2\theta +D_y\sin^2\theta)}p_y  + [{\nu_0\xi_0 q_\bot^2D_y\sin^4\theta \over D_x(D_x\cos^2\theta + D_y\sin^2\theta)}p_y\nonumber \\ &-& i{q_\bot\cos\theta\sin^2\theta(\nu_2\xi_0-\nu_0\xi_2)p_y \over D_x\cos^2\theta + D_y\sin^2\theta}]\Delta\mu + i\xi_Aq_\bot\cos\theta p_y.
\label{pysym3}
\eea
Evidently, the solution for Eq.~(\ref{pysym3}) admits underdamped waves with an anisotropic wavespeed $C(q,\theta)$:
\beq
C(q,\theta)= \frac{q\cos\theta\sin^2\theta(\nu_2\xi_0-\nu_0\xi_2)\Delta\mu}{D_x\cos^2\theta + D_y\sin^2\theta} + \xi_Aq_\bot\cos\theta.
\label{wavespeed}
\eeq
Two situation may arise: (i) when the contribution to the propagating mode from the self-advecting term has the same sign as the one coming from the active terms (top figure in Fig.~\ref{wave}) or (ii) when they have different signs (bottom figure). In general, the wavespeed is anisotropic.
Thus the wavespeed is zero at $\theta = \pi/2$ and nonzero elsewhere in this range of angles with a value that depends on $\theta$ explicitly. Further, by switching the signature of the product $(\nu_2\xi_0-\nu_0\xi_2)\Delta\mu$ the direction of propagation can be reversed.
Writing the solution as $p_y (t)\sim\exp [\lambda(q,\theta)t]$ we find for the eigenvalue $\lambda$ as a function
of $q$ and $\theta$
\begin{eqnarray}
\lambda(q,\theta)&=& -{\kappa q_\bot^2 \over \gamma} - {\kappa\nu_2^2q_\bot^2D_x\cos^2\theta -\kappa\nu_0^2q_\bot^4D_y\sin^4\theta\over D_y(D_x\cos^2\theta +D_y\sin^2\theta)}  + [{\nu_0\xi_0 q_\bot^2D_y\sin^4\theta \over D_x(D_x\cos^2\theta + D_y\sin^2\theta)} \nonumber \\ &-& i{q\cos\theta\sin^2\theta(\nu_2\xi_0-\nu_0\xi_2) \over D_x\cos^2\theta + D_y\sin^2\theta}]\Delta\mu + i\xi_Aq_\bot\cos\theta.
\label{eigcaseIII}
\end{eqnarray}
In general, the real part of $\lambda$ depends on $q$ and $\theta$, reflecting anisotropic growth or decay.  Depending upon the sign of the $\Delta\mu$-dependent terms, there are instabilities at $O(q^2)$ when
\bea
&&{\nu_0\xi_0D_y\Delta\mu \over D_x(D_x\cos^2\theta + D_y\sin^2\theta)}\geq 0,\nonumber \\
{\rm and} \;\; && |\frac{\nu_0\xi_0D_y\Delta\mu}{D_x(D_x\cos^2\theta + D_y\sin^2\theta)}|\sin^4\theta > |\frac{\kappa}{\gamma}+{\kappa\nu_2^2D_x\cos^2\theta \over D_y(D_x\cos^2\theta + D_y\sin^2\theta)}|,
\eea
where the equality sign determines the threshold of instability characterised by a critical $\Delta\mu_c$, such that when $\Delta\mu$ exceeds $\Delta\mu_c$ and the above inequality is met, the system becomes unstable.
Since the phenomenological coefficients $\xi_0$ and $\xi_2$ are in principle independent, such instabilities and underdamped waves can occur
independently or together. The finite wavevector instability  of $O(q^2)$ disappears at $\theta = 0$. At a general angle, instability may be present, depending upon the signs of the relevant coefficients. Therefore, as in {\em System I}, complicated patterns are likely to emerge out of these instabilities. The schematic form of the $q$-dependences of the unstable eigenvalue for an arbitrary $\theta$ is similar to Fig.~\ref{fig1a}: There is a $\theta$-dependent $q_{max}$ at which $\lambda$ is maximum; the maximum instability occurs as this wavevector. We find
\beq
q_{max}= \pm \left[ {\xi_0\Delta\mu \over 2\nu_0\kappa} - {\nu_2^2D_x\cos^2\theta \over 2\nu_0^2D_y\sin^4\theta} - {D_x(D_x\cos^2\theta + D_y\sin^2\theta) \over 2\gamma\nu_0^2D_y\sin^4\theta}\right]^{1/2}
\eeq
Schematic plots of $q_{max}$ versus $\theta$ for two different values of $\Delta\mu$ are shown in Fig.~\ref{q3}. We clearly see (i) $q_{max}$ depends strongly on $\theta$, an illustration of the ensuing anisotropic pattern and (ii) for larger $\Delta\mu (\Delta\mu_{II}>\Delta\mu_I)$, $q_{max}$ is larger, implying that as $\Delta\mu$ rises patterns become denser in the real space. The differences in the dependences of $q_{max}$ on $\theta$ indicate the differences in the generated patterns between {\em System I} and {\em System III}.
Thus, in the real space a direction-dependent anisotropic pattern will emerge with a periodicity $\sim 1/q_{max}$.
A contour plot of eigenvalue $\lambda$ as a function of wavector $q$ and polar angle $\theta$ is shown in Fig. \ref{eigen3} which depicts the stable (negative) and unstable (positive) regions.

The instability that occurs here is for positive values of $\nu_0\xi_0\Delta\mu$. Assuming positive $\nu_0$, this would mean a constant positive active contribution $\xi_0\Delta\mu$ (using $p_x=1$) to $v_x$. In other words, when the active contribution to the velocity is parallel to the direction of the macroscopic order (in this case the $x$-direction) there is instability; when it is anti-parallel to the latter, it stabilises. In terms of the relations between the different coefficients given in Sec.~\ref{configIII} we can get the generalised instability condition as
\bea
{8\kappa (\nu_1-1)^2\cos^2\theta \over W_c^3\pi^2} = -{\kappa \over \gamma_1W_c} - {2W_c^2\Delta\mu(\nu_1+1)\xi'\sin^4\theta \over \eta\pi^4},
\eea
where $W_c$ is the critical thickness in the $z$ direction at which instability sets in. It can have different values in different directions depending upon the values of $\theta$. At $\theta={\pi \over 2}$, the critical thickness is given by
\bea
W_c=|-{\kappa\pi^4\eta \over 2\Delta\mu(\nu_1+1)\xi'}|^{1/3}
\eea
at which instability sets in.

To calculate the correlation function in the incompressible limit we should add a gaussian-distributed thermal noise to Eq.~(\ref{pysym3}) such that it maintains FDT at $\Delta\mu=0$
\begin{equation}
\langle \psi_p({\bf q},t)\psi_p(-{\bf q},0)\rangle = 2K_BT\left[\frac{1}{\gamma}+\frac{\nu_2^2D_x}{D_y}\frac{q_x^2}{D_xq_x^2+D_yq_y^2}
+\frac{\nu_0^2D_y}{D_x}\frac{q_y^4}{D_xq_x^2+D_yq_y^2} \right],
\end{equation}
Assuming small departures from equilibrium we continue to use the same thermal noise even when $\Delta\mu\neq 0$.
This leads us to auto correlation function $\langle p_y^2(x,t)\rangle$
\bea
\langle p_y^2(x,t)\rangle
&=& k_BT\int {q_\bot dq_\bot \over 2\pi}{d\theta \over 2\pi}\left[ {1 \over \gamma} + {\nu_2^2D_x\cos^2\theta \over D_y(D_x\cos^2\theta + D_y\sin^2\theta)} + {\nu_0^2D_y\sin^4\theta q_\bot^2 \over D_x(D_x\cos^2\theta + D_y\sin^2\theta)}\right] \nonumber \\
&&[\kappa q_\bot^2\{ {1 \over \gamma} + {\nu_2^2D_x\cos^2\theta \over D_y(D_x\cos^2\theta + D_y\sin^2\theta)} + {\nu_0^2D_y\sin^4\theta q_\bot^2 \over D_x(D_x\cos^2\theta + D_y\sin^2\theta)}\} \nonumber \\
&& - \Delta\mu{\nu_0\xi_0D_y\sin^4\theta q_\bot^2 \over D_x(D_x\cos^2\theta + D_y\sin^2\theta)}]^{-1}.
\eea
Now if
\bea
[\kappa q_\bot^2\{ {1 \over \gamma} + {\nu_2^2D_x\cos^2\theta \over D_y(D_x\cos^2\theta + D_y\sin^2\theta)}\} - \Delta\mu{\nu_0\xi_0D_y\sin^4\theta q_\bot^2 \over D_x(D_x\cos^2\theta + D_y\sin^2\theta)}]>0,
\eea
for all $\theta$, the system is stable and $\langle p_y^2(x,t)\rangle$ has a logarithmic infrared divergence. 

The diffusion coefficient can be calculated from the velocity autocorrelation function.
The equation for  ${\bf v}$ in the incompressible limit becomes
\bea
v_x &=& [i(\xi_0\Delta\mu q_yP_{xx} - \kappa\nu_0 q_yq_\bot^2P_{xx}) - \kappa\nu_2P_{xy}q_\bot^2 + \xi_2\Delta\mu P_{xy}]p_y + P_{xj}f_j, \nonumber \\
v_y &=& [i(\xi_0\Delta\mu q_yP_{xy} - \kappa\nu_0 q_yq_\bot^2 P_{xy}) + \xi_2\Delta\mu P_{yy} - \nu_2\kappa q_\bot^2 P_{yy}]p_y + P_{yj}f_j,
\eea
where we have added a thermal noise $f_i$ with correlations given by Eq.~(\ref{vncorr}),
and $P_{ij}=(\delta_{ij} - {q_iq_j \over q_\bot^2})$ is the transverse projection operator. We find that the diffusion coefficient $D_{xx}$ has infra-red divergent parts which depend on $\Delta\mu$. Considering only the most infra-red divergent part, we obtain
\bea
D_{xx} &=& {1 \over 2}\int {d^2q_\bot \over (2\pi)^2}\langle v_x(q_\bot,\omega=0)v_x(-q_\bot,\omega=0)\rangle \nonumber \\
&=& k_BT\int {d^2q_\bot \over (2\pi)^2}[(\xi_0\Delta\mu q_yP_{xx})^2 + (\xi_2\Delta\mu P_{xy})^2]\left[{1 \over \gamma} + {\nu_2^2\cos^2\theta \over D} + {\nu_0^2q_\bot^2\sin^4\theta \over D}\right]\nonumber \\
&&\times [\{\frac{\Delta\mu q_\bot \cos\theta\sin^2\theta(\nu_2\xi_0 - \nu_0\xi_2)} { D} - \xi_Aq_\bot\cos\theta \}^2 + \{ \kappa q_\bot^2({1 \over \gamma} + {\nu_2^2\cos^2\theta \over D} + {\nu_0^2\sin^4\theta q_\bot^2 \over D})\nonumber \\
&& - {\Delta\mu \nu_0\xi_0\sin^4\theta q_\bot^2 \over D}\}^2]^{-1}.
\eea
Therefore, the diffusion coefficient diverges as $\ln L$, where $L$ is the lateral system size. Thus we find diverging diffusion coefficient in {\em generic polar systems}, in contrast to System I, where such divergences are found only for nematic order. No such divergences exist in System I with polar order. This is directly related to the presence of the active external forces. Nonlinear terms, not considered here, are likely to modify the $L$-dependences \cite{toner}. Thus, $D_{xx}$ diverges as $L^2$ where $L$ is the linear size of the system (in the $XY$-plane). Further, near the threshold of linear instability, $D_{xx}\sim
(\mu_c -\mu)^{-2}$ where $\mu_c$ the critical activity at the onset of linear instability in {\em System III}.

\section{Summary and outlook}
\label{conclu}

The aim of this article is to set up a stochastically driven generalised hydrodynamic theory of thin confined active gels as a $2d$ generic model for cortical actin in eukaryotic cells. The presence of a substrate (the bulk of the cell) breaks Galilean invariance of the system. We show how {\em effective} $2d$ descriptions emerge for thin confined $3d$ active gel systems by integrating the thickness dependences. This allow us to write down linear constitutive relations between them, subject to spatial symmetries and conservation laws. These include terms linear in $\Delta\mu$, representing small deviations from equilibrium. We show that the pairs $(v_i, \dot p_i)$ and $(F_i=-\partial_i \Pi, h_{\perp i})$ are good candidates for thermodynamic fluxes and forces in the present problem. In particular, we consider three different cases - (i) in-plane nematic or polar order without any external force, (ii) polar order normal to the plane without any external force and (iii) in-plane polar order with external forces. The ensuing  equations of motion of the relevant slow variables display linear instabilities at finite wavevectors for certain values and signs of the activity. In the next step, following the principles of Onsager reciprocity theorem, we add thermal noises such that the FDT is maintained in the absence of any activity. For finite activity, the FDT is manifestly broken as is expected. The noisy equations of motion may display linear instability, depending upon the sign and magnitude of $\Delta\mu$. For the stable case, we use them to calculate the correlation functions of the appropriate variables in the nonequilibrium steady state. In addition, we calculate the diffusion coefficients of tagged particles for System I and III, and show that they diverge for a given strength of the activity. In the case when there are no linear instabilities, we show that the equal-time density autocorrelation function show giant fluctuations under generic conditions. Our descriptions are expected to be relevant for rigid biofilms \cite{wiki} made of bacteria colonies on surfaces, e.g., {\em Pseudomonus aeruginusa} \cite{pseudo} and microtubule arrays in eukaryotic cells and actin motility assays. Equations similar to those presented here have been discussed elsewhere in various contexts, see, e.g., Ref.~\cite{simha,toner,gruler}. Our presentations here complement the existing work. Controlled cell biology experiments are needed to test the predictions from our models. Numerical simulations may be helpful in this regard. Giant fluctuations in active particle systems have already been observed in numerical experiments \cite{chate}. We look forward to further detailed numerical work using these equations with more realistic situations in the future.
%

Our $2d$ equations of motion are linear as we have neglected the nonlinear terms. In a more complete theoretical framework, nonlinear terms should be included. They are expected to be responsible in arresting the linear instabilities and ensuring transition to a non-trivial steady state (may be a nonuniform or a flowing steady state). The statistical properties in such steady states are expected to be much richer and complex in nature \cite{luca}. Even in the case where there is no instability, nonlinear terms may modify the scaling properties of the correlation functions obtained at the linear level \cite{toner,sradha}. Such issues are intimately connected to the general question of the nature of order in a $2d$ driven nonequilibrium system with broken continuous symmetries. Further work is in progress in this direction \cite{niladri2}.
\section{Acknowledgement}
We thank J.-F Joanny and S. Ramaswamy for many fruitful discussions and critical comments on the manuscript. One of us (AB) gratefully acknowledges partial financial support in the form of the Max-Planck Partner Group at the Saha Institute of Nuclear Physics, Kolkata funded jointly by the Max-Planck-Gesellschaft (Germany) and Department of Science and Technology (India).

\section{Appendix A}
\label{appen1}

Here we derive the effective $2d$ constitutive equations for the nematic and polar sample in System I directly by using symmetry arguments in $2d$. As discussed above, the $2d$ sample is invariant under $y \rightarrow -y$ as nothing distinguishes $y$ from $-y$ direction. However, due to macroscopic ordering along the $x$-direction ($p_x=1$) there is no symmetry under $x \rightarrow -x$. We first consider nematic order and hence ${\bf p}\rightarrow -{\bf p}$ symmetry is present. In the absence of any external forces, we have ${\bf v}\propto \nabla(\cdot)$, such that ${\bf v}({\bf q=0},t)=0$ where ${\bf q}$ is a Fourier wavevector. Using these arguments we can write down the constitutive relations for $v_x$ and $v_y$ as linear functions of the relevant fluxes $F_x$, $F_y$, $h_x$ and $h_y$. In order to satisfy the required symmetry dictated tensorial nature of the constitutive relations, the Onsager matrix is to be constructed out of $\bf p$ (with $p^2=1$ here by choice) or $\partial_x,\,\partial_y$.  In general, the constitutive relations for $v_i$ have the form
\bea
v_i=DF_i+ \nu_0\partial_j (p_i h_j) +\nu_2\partial_j (p_j h_i) +\overline\nu\partial_i ({\bf p\cdot h}). \label{case1onsv}
\eea
where $\nu_0,\nu_2,\overline\nu_2$ are dissipative coefficients coupling the flow with the conjugate orienting field $\bf h_\perp$.
Next we include a nonequilibrium drive, linear in $\Delta\mu$. Microscopically, the conversion of ATP to ADP acts a chemical fuel creating an {\em active stress} (or {\em active force}) which acts on the flow field over and above the equilibrium generalised forces. In addition to the equilibrium terms, for an active gel, there will be active contributions to $v_i$. The most relevant form of such a contribution to $v_i$ is $\sim \Delta\mu \partial_j( p_i p_j)$, where $\Delta\mu p_ip_j$ is an active stress. Therefore the equations for $v_x$ and $v_y$ take the form (up to the lowest order of gradients and linear in $\Delta\mu$)
\bea
v_x &=& DF_x + \nu_0 \partial_y (p_x h_{\perp y}) +\nu_x\partial_x(p_x h_{\perp x}) +\nu_2\partial_y(p_x h_{\perp y})+\xi_0 \Delta\mu \partial_y(p_xp_y),\label{case1vxA} \\
v_y &=& DF_y + \nu_2\partial_x (p_x h_{\perp y}) + \overline\nu\partial_y (p_x h_{\perp x})+ \nu_0 \partial_x(p_y h_{\perp x})
+ \xi_2\Delta\mu \partial_x(p_xp_y),\label{case1vyA}
\eea
where we have retained up the the linear order in $p_y$, $F_i=-\nabla_i\Pi$, $\nu_x=\nu_0+\nu_2+\overline\nu_2$, $\xi_0$ and $\xi_2$ are the coupling constants coupling activity $\Delta\mu$ to the flow fields. We have retained terms up to $O(p_y)$ in smallness in the active terms, and hence, possible active contributions like $\Delta\mu\partial_x (p_y^2)$ to $v_x$ and $\Delta\mu\partial_y (p_y^2)$ to $v_y$ are ignored to this order. Further, an active contribution of the form $\Delta\mu\partial_x (p_x^2)$ yields zero to leading order in smallness since $p_x=1$ to that order. Unsurprisingly, active terms in Eqs.~(\ref{case1vxA}-\ref{case1vyA}) respect the spatial symmetries of the constitutive relation (\ref{case1onsv}) in equilibrium. We have used same coefficients $\nu_0,\,\nu_2,\,\xi_0,\,\xi_2$ as in Sec.~\ref{config1} in anticipation of obtaining equations identical to Eqs.~(\ref{vxIfinal}-\ref{pyIfinal}) here. The above equations (\ref{case1vxA}) and (\ref{case1vyA}) generalise the Darcy's law for a simple fluid to an active (nonequilibrium) gel.

In general in equilibrium the dynamics of $p_x,\,p_y$ have a term representing relaxational dynamics towards local equilibrium and  terms which couple it to ${\bf v}$. In this case the equation for $p_i$ is
\bea
\frac{\partial p_i}{\partial t} &=& \frac{h_{\perp i}}{\gamma} + \psi_i (x,y),
\label{case1pi}
\eea
where $\gamma$ is the rotational viscosity. The explicit form of the vector-valued function $\psi_i$  can be found out using the following arguments in equilibrium ($\Delta\mu =0$): Since in Eqs.~(\ref{case1vxA}) and (\ref{case1vyA}) for ${\bf v}$ we have  terms proportional to $\nabla h_{\perp x}$ and $\nabla h_{\perp y}$, we expect on symmetry ground (Onsager principle) only gradients of ${\bf v}$ will appear in the $p_x$ and $p_y$ equations. The signs and the coefficients are to be chosen in such a way that the Onsagers reciprocity theorem holds in equilibrium. Thus Eq.~(\ref{case1pi}) generally takes the form ($\Delta\mu =0$), with $i=x,y$
\bea
\frac{\partial p_x}{\partial t}&=&\frac{h_{\perp x}}{\gamma} - \alpha_1 p_x\partial_x v_x - \alpha_3 p_x\partial_y v_y,\label{case1px2},\\
{\partial p_y \over \partial t} &=& {h_{\perp y} \over \gamma} - \alpha_0 p_x\partial_yv_x - \alpha_2p_x\partial_xv_y, \label{case1py2}
\eea
where the coefficients $\alpha_1,\,\alpha_3,\,\alpha_0$ and $\alpha_2$ are to be determined using the Onsager symmetry (reciprocity theorem; see below). When $\Delta\mu\neq 0$, explicit dependence of $v_i$ on $\Delta\mu$ generates activity-dependent terms in the equations for $p_i$. One may further add an active term of the form $\lambda_1\Delta\mu p_x$ in Eq.~\ref{case1px2} to obtain an explicit equation for $p_x$ by substituting for $v_x$ and $v_y$ from Eqs.~(\ref{case1vxA}) and (\ref{case1vyA}).
\bea
\frac{\partial p_x}{\partial t} &=& \frac{h_{\perp x}}{\gamma} - \alpha_1\left[\nu_xp_x\partial_x^2 (p_x h_{\perp x}) +\nu_0p_x\partial_x\partial_y (p_x h_{\perp y})\right]-\alpha_3 p_x\left[\nu_2\partial_x\partial_y (p_x h_{\perp y}) + \overline\nu\partial_y^2 (p_xh_{\perp x})\right] \nonumber \\ &-&\alpha_1Dp_x\partial_x F_x-\alpha_3 Dp_x\partial_y F_y +\lambda_1\Delta\mu p_x -\alpha_1\xi_0p_x\Delta\mu\partial_x \partial_y(p_x p_y)-
\alpha_3\xi_2p_x\Delta\mu\partial_x^2 (p_x p_y).
\label{pxIdelmu}
\eea
However, we do not add any active term proportional to $\Delta\mu p_y$ directly in the $p_y$ equation, since $p_y$ is the transverse fluctuation and is a broken symmetry variable \cite{footnote}.
Substituting Eqs. (\ref{case1vxA}) and (\ref{case1vyA}) in Eq. (\ref{case1py2}) we get the equation for orientational field to the linear order in gradients as
\bea
{\partial p_y \over \partial t} &=& \left[{h_{\perp y} \over \gamma} - p_x\nu_0 p_x\partial_y^2 (p_x h_{\perp y} - p_x\alpha_2\nu_2\partial_x^2(p_xh_{\perp y})\right]- \alpha_0 Dp_x \partial_y F_x -\alpha_2 Dp_x\partial_x F_y \nonumber \\
&&  -\alpha_0\xi_0\Delta\mu p_x\partial_y^2 (p_xp_y) - \alpha_2\xi_2\Delta\mu p_x\partial_x^2(p_xp_y) -p_x[\alpha_0\nu_x +\alpha_2\overline\nu]
\partial_x\partial_y (p_x h_{\perp x}). \label{case1py3}
\eea

We use Onsager's reciprocity relation in equilibrium ($\Delta\mu=0$) to obtain
\bea
\alpha_0={\nu_0 \over D},\;\;
\alpha_2= {\nu_2 \over D},\;\;\alpha_1=\frac{\nu_0+\nu_2+\overline\nu} {D},\;\;\alpha_3={{\overline\nu}\over D}.
\label{alphaI}
\eea

In a renormalised theory coefficients $\alpha_0$ and $\alpha_2$ should generally acquire a $\Delta\mu$-dependence. However, restricting ourselves within the scope of a linear flux-force relationship we continue to use the same $\alpha_0,\,\alpha_1,\,\alpha_2$ and $\alpha_3$ as in Eqs.~(\ref{alphaI}) for small $\Delta\mu$ which yield the correct equilibrium limit.
The active terms affect $h_{\perp x}$ which may be determined as before by setting $p_x=1$ in Eq.~(\ref{pxIdelmu}) to the leading order in $p_y$. We obtain
\bea
h_{\perp x}=-\gamma\left[\lambda_1\Delta\mu+\frac{\nu_x\nu_0}{D}\partial_x\partial_y  h_{\perp y}+\frac{\nu_2\nu_y}{D}\partial_x\partial_y  h_{\perp y}-\nu_x\partial_x F_x -\nu_y\partial_y F_y\right].
\eea
Substituting $h_{\perp x}$ in Eqs.~(\ref{case1vxA}) and (\ref{case1vyA}), we note that the only effect of $h_{\perp x}$ is to shift the coefficients $\xi_0$ and $\xi_2$ of the active terms by amount $\lambda_1$. Thus the {\em nonequilibrium} version of the Darcy's law in this case become
\bea
v_x &=& DF_x + \nu_0 \partial_y (p_x h_{\perp y}) +\xi_0 \Delta\mu \partial_y(p_xp_y),\label{case1vx}\\
v_y &=& DF_y + \nu_2\partial_x (p_x h_{\perp y}) +\xi_2 \Delta\mu \partial_x(p_xp_y)\label{case1vy},
\eea
where coefficients $\xi_0$ and $\xi_2$ are to be understood as {\em effective coefficients} after absorbing the contributions coming from elimination of $h_{\perp x}$.  One may, however, add new relevant active terms bilinear in ${\bf p}$ and linear in $\nabla$ and $\Delta\mu$ in Eq.~(\ref{case1py2}); such terms however break the nematic symmetry considered here. We consider the effects of such terms separately below.
Finally, putting these values of $\alpha_0$ and $\alpha_2$ and eliminating $h_{\perp x}$, Eq.~(\ref{case1py3}) becomes
\bea
{\partial p_y \over \partial t} &=& \left[{h_{\perp y} \over \gamma} - {\nu_0^2 \over D}p_x\partial_y^2(p_x h_{\perp y}) - {\nu_2^2 \over D}p_x\partial_x^2(p_x h_{\perp y})\right] -\nu_0p_x\partial_y F_x -\nu_2 p_x\partial_x F_y \nonumber \\
&&  -{\nu_0 \over D}\xi_1\Delta\mu p_x\partial_y^2(p_xp_y) - {\nu_2 \over D}p_x\xi_2\Delta\mu\partial_x^2(p_xp_y).\label{case1py}
\eea

It is evident that the Eqs.~(\ref{case1vx}), (\ref{case1vy}) and (\ref{case1py}) are invariant under
\bea
&(i)& y\rightarrow -y\;\; ({\rm no\;\;distinction\;\;between}\, y\, {\rm and}\, -y), \nonumber \\
&(ii)& {\bf p}\rightarrow -{\bf p} \;\; ({\rm nematic\;\; symmetry}).
\eea
Further note that the coefficients $D,\nu_0,\,\nu_2$ and $\gamma$ form the Onsager coefficient matrix which is symmetric, as is expected.  The signatures of these coefficients are determined by the positivity of dissipation. Substituting Eqs.~(\ref{case1vx},\,\ref{case1vy}) and (\ref{case1py}) in Eq.~(\ref{freenderi}) and demanding that each term makes a positive contribution to dissipation separately, we find $D>0,\,\gamma>0,\,\nu_0>0$ and $\nu_2>0$.

The above equations (\ref{case1vx}), (\ref{case1vy}) and (\ref{case1py}) exhibit the nematic order. In order to introduce polar order this inversion symmetry must be broken. Hence we add a term $\xi_A ({\bf p}\cdot \nabla_\bot){\bf p}$ to the equation for $\dot{p_y}$. Since $p_x=1$, the only term, which survives upon linearising about the chosen reference state, is $\xi_A {\partial p_y \over \partial x}$. Here, $\xi_A$ is an active coefficient, proportional to $\Delta\mu$. Hence the equation for $\dot{p_y}$ having polar order is
\bea
\dot{p_y}&=& \left[{h_{\perp y} \over \gamma} - {\nu_0^2 \over D}p_x\partial_y^2 (p_xh_{\perp y})- {\nu_2^2 \over D}p_x\partial_x^2(p_xh_{\perp y})\right]-\nu_0p_x\partial_y F_x -\nu_2 p_x\partial_x F_y \nonumber \\
&&  -{\nu_0 \over D}\xi_0\Delta\mu p_x\partial_y^2 (p_xp_y) - {\nu_2 \over D}\xi_2\Delta\mu p_x\partial_x^2 (p_xp_y) + \xi_Ap_x{\partial p_y \over \partial x}.
\label{polarpcaseIappen}
\eea
Note that Eq.~(\ref{polarpcaseIappen}) breaks the invariance under $\bf p\rightarrow -p$ and (separately) under $\bf r\rightarrow -r$. The lack of the former makes the system polar; the lack of the latter leads to propagating wave-like excitations, as we will see below. Finally, using $p_x=1$ we obtain Eqs.~(\ref{vxIfinal}-\ref{pyIfinal}).

\section{Appendix B}
\label{appen2}


Here we derive the $2d$ equations directly for a sample with external forces. In this sample the total generalised forces, due to the presence of external forcs, is not zero and must be proportional to either ${\bf h}_\bot ({q=0},t)$ or ${\bf p}({\bf q=0},t)\Delta\mu$ (${\bf q}$ is a fourier wavevector). As the system is polar we do not have ${\bf p}\rightarrow -{\bf p}$ symmetry. We further consider planar alignment, i.e., in this case we have $\langle {\bf p}\rangle=p_0{\bf e_x}={\bf e_x}\neq 0$ in the plane. Hence, there is no $x\rightarrow -x$ symmetry. However, the system is still invariant under $y\rightarrow -y$. We continue to impose $p^2=1$. In addition we consider departure from equilibrium to the linear order in $\Delta\mu$. We consider a situation where the system experiences external forces (e.g., forces imparted to the actin filaments by the immobile molecular motors grafted on the confining substrate in actin motility assays) which, on symmetry ground, should be proportional to $h_i$ or $\Delta\mu p_i$. From the invariance properties discussed above, we can write the constitutive relations for the velocity fields such that the Onsager reciprocity principle holds in equilibrium ($\Delta\mu=0$)
\bea
v_x &=& DF_x + \nu_0 \partial_y (p_x h_{\perp y}) + \nu_2 h_{\perp x} +\xi_0\Delta\mu \partial_y(p_xp_y)+\xi_0\Delta\mu \partial_y(p_xp_y), \label{case3vx1} \\
v_y &=& DF_y + \nu_2h_{\perp y} + \xi_2\Delta\mu p_y, \label{case3vy1}.
\eea
where we have retained terms with leading order spatial gradients, up to linear order in $p_y$ and active terms up to linear order in $\Delta\mu$ having same symmetries as the equilibrium terms. Coefficient $D>0$ is the inverse friction coefficient, and $\nu_0$, $\nu_2$ and $\nu_y$ are dissipative cross-coupling coefficients, coupling flows with the local orientation, $\xi_0$ and $\xi_2$ are coefficients of the active terms.

Equilibrium dynamics of the orientational field comes from its relaxation to local equilibrium and its coupling to flows. Hence the general form of the orientational field can be written as (to leading order in gradients)
\bea
\dot{p_x}&=&\frac{h_{\perp x}}{\gamma} + \alpha_x v_x +\alpha_yp_x \partial_y v_y,\label{px3inter}\\
\dot{p_y}
 &=& {h_{\perp y} \over \gamma} + \alpha_0v_y + \alpha_2 p_x\partial_yv_x.
\label{interIII}
\eea
where $\alpha_x,\,\alpha_0,\,\alpha_2$ and $\alpha_y$ are coupling constants coupling the orientation field $p_x,\,p_y$ to the flows, all of which can be determined by using the Onsager reciprocity theorem in equilibrium, i.e., $\Delta\mu=0$. When the active terms in Eqs.~(\ref{case3vx1}) and (\ref{case3vy1}) are non-zero, explicit dependences of $v_x$ and $v_y$ generate $\Delta\mu$-dependences of $p_x$ and $p_y$. In addition we add an active term $\lambda_1\Delta\mu p_x$ in the $p_x$-equation. However, a similar term does not exist in the dynamical equation of $p_y$, as to the lowest order $p_y$ is the transverse mode.
Put together everything, we get (to the leading order in gradients)
\bea
\dot{p_x}&=&\frac{h_{\perp x}}{\gamma} +\alpha_x [DF_x + \nu_2 h_{\perp x}+
\nu_0 \partial_y (p_x h_{\perp y})] +\alpha_y p_x \partial_y [DF_y +\nu_2h_{\perp y} + \nu_y (p_x h_{\perp x})] \nonumber \\&+&\lambda_1\Delta\mu p_x
+\alpha_x\xi_0\Delta\mu \partial_y (p_x p_y)+\alpha_y \xi_2\Delta\mu p_x \partial_y p_y,\label{case3px}\\
\dot{p_y} &=& [{1 \over \gamma} + \alpha_0\nu_2 - \alpha_2\nu_0\partial_y^2]h_{\perp y} + \alpha_0DF_y  + \alpha_2D\partial_yF_x + \alpha_2\nu_2 p_x \partial_y h_{\perp x}\nonumber \\ &+&\alpha_2\xi_0\Delta\mu p_x\partial_y^2 p_y. \label{case3py}
\eea
The condition $p_x=1$ yields $h_{\perp x}$, which upon substituting in Eqs.~(\ref{case3vx1}) and (\ref{case3vy1}) give (for $\Delta\mu=0$):
\bea
v_x &=&D\left[1  -\frac{\nu_2\alpha_x}{\frac{1}{\gamma} +\alpha_x \nu_2}\right]F_x + \nu_0  \left[1- \frac{\nu_2\alpha_x}{\frac{1}{\gamma}+\alpha_x\nu_2} \right]\partial_y (p_x h_{\perp y}),\\
v_y &=& D F_y +\nu_2 h_{\perp y}.
\eea
Thus the {\em effective} friction coefficients are no longer isotropic.
Further, Onsager reciprocity theorem requires
\beq
\alpha_0 = {\nu_2 \over D}, \;\;
\alpha_2 = -{\nu_0 \over D},\;\;\alpha_x = \frac{\nu_2}{D},\,\alpha_y=-\frac{\nu_0}{D}.
\eeq
Defining $D_x=D^2/(D+\nu_2^2 \gamma),\,D_y=D$ and $\nu_0'=\nu_0D/(1+\gamma\nu_2^2)$, we write
\bea
v_x &=& D_x F_x + \nu_0'\partial_y (p_x h_{\perp y}),\\
v_y&=& D_y F_y + \nu_2 h_{\perp y},
\eea
Evidently, the fiction coefficients are now anisotropic and $D_x<D_y$. This is in contrast to the nematic sample or System I, which we considered above, where fiction coefficients are isotropic. For the purpose of brevity, we represent $\nu_0'$ by $\nu_0$.
As in {\em Appendix A}, we continue to use the same equilibrium values for $\alpha_0,\,\alpha_2$ and $\alpha_x$ even when $\Delta\mu\neq 0$. Then using their values together with $h_{\perp x}$ and setting $p_x=1$ we then obtain the  constitutive relations (\ref{case3vx} - \ref{case3ons}).


\end{document}